\documentclass[a4paper,11pt]{article}
\usepackage{amsmath,amssymb,bm}
\usepackage{titlesec}
\titleformat{\paragraph}[runin]{\normalfont\itshape}{\theparagraph.}{.3em}{}[.]\titlespacing{\paragraph}{0pt}{1ex plus .1ex minus .2ex}{.5em}
\usepackage[letterpaper, hmargin=1in, top=1in, bottom=1.2in, footskip=0.6in]{geometry}
\usepackage[utf8]{inputenc}
\usepackage{lmodern}
\usepackage{mathtools}
\usepackage{dsfont}
\pdfoutput=1
\usepackage[T1]{fontenc}
\usepackage[english]{babel}
\usepackage{graphicx} 
\usepackage{booktabs} 
\usepackage{color}
\definecolor{aquamarine}{rgb}{0.5, 1.0, 0.83}
\definecolor{ao(english)}{rgb}{0.0, 0.5, 0.0}
\definecolor{armygreen}{rgb}{0.29, 0.33, 0.13}
\definecolor{awesome}{rgb}{1.0, 0.13, 0.32}
\definecolor{ballblue}{rgb}{0.13, 0.67, 0.8}
\definecolor{bittersweet}{rgb}{1.0, 0.44, 0.37}
\definecolor{blue}{rgb}{0.0, 0.0, 1.0}
\definecolor{brinkpink}{rgb}{0.98, 0.38, 0.5}
\definecolor{ballblue}{rgb}{0.13, 0.67, 0.8}
\definecolor{brightturquoise}{rgb}{0.03, 0.91, 0.87}
\definecolor{blue-green}{rgb}{0.0, 0.87, 0.87}
\definecolor{caribbeangreen}{rgb}{0.0, 0.8, 0.6}
\definecolor{cyan}{rgb}{0.0, 1.0, 1.0}
\definecolor{amber(sae/ece)}{rgb}{1.0, 0.49, 0.0}
\graphicspath{ {images/} }

\author{J\"{u}rg Fr\"{o}hlich\footnote{Institute of Theoretical Physics, $ETH$ Zurich, 8093 Zurich, Switzerland; Email: juerg@phys.ethz.ch} 
\quad Zhou Gang\footnote{Department of Mathematics and Statistics, Binghamton University, Binghamton, NY 13902, USA; Email: gangzhou@binghamton.edu}\quad Alessandro Pizzo\footnote{Dipartimento di Matematica, Universit\`a di Roma ``Tor Vergata", 
00133 Roma, Italy; Email: pizzo@mat.uniroma2.it}}

\title{A Theory of Quantum Jumps}

\begin{document}

\maketitle

\begin{abstract}
Using the principles of the $ETH$ - Approach to Quantum Mechanics we study fluorescence and the phenomenon 
of ``quantum jumps'' in idealized models of atoms coupled to the quantized electromagnetic field. In a limiting regime 
where the orbital motion of the atoms is neglected and the velocity of light tends to $\infty$ we derive explicit non-linear 
stochastic differential equations describing the effective time evolution of states of individual atoms. These equations 
give rise to a measure on state-trajectories with quantum jumps which is a quantum-mechanical analogue of the Wiener 
measure of Brownian motion. Our results amount to a derivation of the fundamental randomness in the 
quantum-mechanical description of microscopic systems from basic principles in the context of some simple models.
\end{abstract}

\tableofcontents

\section{Introduction: Stochasticity in Quantum Mechanics}\label{Intro}
It has been widely appreciated ever since the discovery of Quantum Mechanics (QM) that the quantum-mechanical 
time evolution of states of \textit{individual} physical systems appears to be \textit{non-linear} and \textit{stochastic}, 
while the time evolution of averages of states over a large ensemble of identical, identically prepared systems is \textit{linear} 
and \textit{deterministic}. The latter evolution is described by a Schr\"odinger - von Neumann equation, or by a Lindblad-type
equation (or generalizations thereof).

The stochastic nature of the quantum-mechanical evolution of individual systems was first recognized by \textit{Einstein}
in his paper \cite{Einstein} on spontaneous and induced emission and absorption of photons by atoms. In his famous book 
\cite{Heisenberg}, \textit{Heisenberg} quotes Einstein as declaring the following:  \textit{``You know  that I suggested the idea 
that the atom \textbf{suddenly} falls from one stationary state down to a stationary state of lower energy by emitting the energy 
difference in the form of a quantum of light''} (i.e., a photon). Einstein understood that the time when such a \textit{``quantum jump''} 
happens and the characteristics of the photon emission accompanying it are \textit{random objects} which cannot be predicted 
with certainty by the theory, implicating that ``quantum jumps'' are stochastic.  After the discovery of matrix mechanics, in 1925, 
the probabilistic nature of quantum theory was emphasized by \textit{Born} in his famous paper \cite{Born} on \textit{``Stossvorg\"ange''}. 
These and other publications gave rise to the metaphor of  ``quantum jumps''\footnote{See remarks in \cite{Heisenberg}; for fairly recent 
results on ``quantum jumps,'' see, e.g., \cite{Zoller, C-T-Zambon} and references given there.} exalted by some of the pioneers of QM, 
but disliked and disbelieved by others, such as Einstein and \textit{Schr\"odinger}. Be that as it may, if one analyzes 
concrete phenomena, such as the flashes of light emitted from a spot in a screen hit by a particle in the famous double-slit 
experiment, the Stern-Gerlach experiment, the fluorescence of atoms, or the radioactive decay of unstable nuclei, one comes to 
the conclusion that the stochastic behavior of microscopic systems is an observational fact established beyond any reasonable 
doubt. A satisfactory completion of Quantum Mechanics ought to account for this behavior.

Unfortunately, there is no consensus as to the origin of randomness in QM; nor is there a consensus concerning the 
precise form of a fundamental \textit{quantum-mechanical law} describing the stochastic time evolution of states of 
\textit{individual} physical systems. To quote \textit{Dirac},  \textit{``it seems clear that the present quantum mechanics 
is not in its final form.''} Thus, one has to try to cast QM in a final form by searching for such a law!

Many theorists are under the spell of one or another variant of the \textit{Copenhagen Interpretation of QM} and
thus adhere to the opinion that the root of the fundamental randomness of quantum-mechanical phenomena 
lies in the \textit{measurement process.} In particular, it is very often claimed that the time evolution of the state of 
an \textit{isolated} system is governed by a \textit{linear, deterministic} Schr\"odinger - von Neumann equation, 
\textit{unless} a measurement is made, in which case the system's state may jump into an eigenstate of the 
physical quantity that is being measured. This jump- or collapse process is \textit{not} described by a linear deterministic 
evolution equation; it must be described by a non-linear stochastic equation. Unfortunately, it is usually left vague what 
a measurement process consists in, what is special about it, and why it interrupts Schr\"odinger evolution of states of the 
\textit{total} system including all measuring devices. 

Elaborating on Born's Rule, postulates concerning the measurement process were proposed by 
von Neumann \cite{vN} and L\"uders \cite{Luders} (among others). Von Neumann's treatment of measurements
involves the idea that \textit{pure states} may evolve into \textit{mixtures,} i.e., that a form of \textit{dissipation}
(of information) is observed when measurements are carried out. This is usually thought to be an effect of 
\textit{``decoherence,''} i.e., of interactions of a system of interest with some ``environment'';  see, e.g., \cite{Hepp, Zurek}. 
Unfortunately, the separation between ``system of interest'' and ``environment'' tends to be fluid and to be 
made in an ad-hoc way. L\"uders' postulate incorporates the idea of \textit{``state reduction,''} or \textit{``wave-function collapse''}
happening in the course of a measurement; but it does not describe any mechanism causing the collapse.\footnote{Mechanisms 
of spontaneous wave-function collapse have been proposed in \cite{GRW} -- however artificial and speculative they may seem.}

The measurement postulates of von Neumann and L\"uders have been revisited and extended by many people. 
For example, there are proposals of \textit{non-linear stochastic evolution equations} describing the evolution of states
of individual systems under ``(continuous) measurements''; see \cite{Barchielli, Gisin, Diosi, Primas, BDH, ...}. 
In derivations of such equations found in the literature, the only principle invoked appears to be that if 
solutions are averaged over the randomness one obtains density matrices evolving according to a \textit{linear 
deterministic} equation, typically a Lindblad equation \cite{Lindblad}.
While some of the stochastic equations proposed in the literature (see, e.g., \cite{Gisin} and references given there)
provide plausible descriptions of the time evolution of states of \textit{individual} systems, some others -- for example 
those describing stochastic evolution driven by Brownian motion -- look rather contrived. What, in our opinion, is missing 
is a clear understanding of the \textit{physical basis} of such equations, i.e., of physically plausible principles 
implying a precise form of these equations. In particular, the idea that the randomness in the evolution of 
individual physical systems is caused by \textit{``measurements''} (whatever this may mean concretely) strikes 
us as highly questionable. Considering processes such as the fluorescence of atoms, the decay of unstable 
states, or structure formation in the early universe, one is led to expect that there must be a 
\textit{fundamental cause of randomness} in the quantum-mechanical evolution of individual systems \textit{not} involving 
interventions of ``observers'' or ``agents'' and unrelated to measurements.

A fundamental mechanism causing randomness in the evolution of states of individual isolated systems and dissipation 
in the evolution of averages of states over large ensembles of identical, identically prepared systems has been 
proposed in \cite{FS, BFS, FGP}. The ideas developed in these papers are known as the \textit{``$ETH$ - Approach to Quantum Mechanics.''} According to this approach, quantized fields describing massless modes, in particular photons, play a 
central role in causing the fundamental randomness of quantum-mechanical evolution. In fact, the role played 
by the electromagnetic field (and, ultimately, also by the gravitational field) in a \textit{physically viable completion} of Quantum 
Mechanics, solving, among other problems, the so-called ``measurement problem,'' appears to be as fundamental as 
the role these fields have played in the foundation of Relativity Theory. Results in \cite{Buchholz} on scattering theory in
quantum electrodynamics have motivated the emphasis put on the role of the quantized electromagnetic field in the 
$ETH$ - Approach -- but this approach really has a broader basis.

\subsection{Purpose and organization of paper}
The purpose of this paper is to illustrate some basic features of the $ETH$ - Approach to Quantum Mechanics 
in the context of idealized models of fluorescence (see, e.g., \cite{Pomeau} and references given there). These models describe 
physical systems consisting of atoms with finitely many internal energy levels interacting with the quantized electromagnetic field. 
A much studied model of such systems is the \textit{Pauli-Fierz model} of quantum electrodynamics with non-relativistic
matter; see, e.g., \cite{C-T}.  Among phenomena in quantum optics that have been studied mathematically on the basis 
of the Pauli-Fierz model are atomic resonances, spontaneous emission of photons by atoms and Rayleigh scattering. 
Examples of rigorous results on the Pauli-Fierz model are contained in \cite{BaFSi, BFP, FGS}; but there are many further 
important contributions. Unfortunately, it appears that the Pauli-Fierz model only describes the behavior of \textit{ensembles} 
of many identical, identically prepared systems of atoms; but it fails to describe the stochastic evolution of \textit{individual} 
systems.\footnote{It is likely that the Pauli-Fierz model does not satisfy all the premises of the $ETH$ - Approach and hence will not be
studied in this paper.} Motivated by the desire to come up with very explicit, mathematically precise results on the stochastic 
evolution of states of individual systems of atoms interacting with the quantized electromagnetic field, in particular on spontaneous 
emission of photons, we will consider the limiting regime where the orbital motion of the atoms is neglected and the velocity of light, 
$c$, tends to $\infty$. We will study the spontaneous emission of photons by atoms in two situations:\\ 
(i) Photons emitted by an atom escape and are never recorded; and\\ 
(ii) photons emitted by an atom are recorded by triggering the firing of a photomultiplier.\\
These processes will be analyzed on the basis of the $ETH$ - Approach, making use of the so-called 
\textit{``Principle of Diminishing Potentialities''} and the \textit{``State-Reduction Postulate''}; see \cite{FS, BFS, FGP, FP}. 
In the context of the models studied in this paper, these two building blocks of the $ETH$ - Approach imply 
specific non-linear stochastic equations for the time evolution of states of  individual atoms, as we will show in Sects.~\ref{Unraveling Lindblad},
\ref{Q-Poisson} and \ref{Example}.
It will turn out that these equations depend on whether photons emitted by an atom are recorded, 
or not. If these photons are \textit{not} recorded we recover effective evolution equations for the states of individual 
atoms that have already been considered in, e.g., \cite{Gisin}; if the photons \textit{are} recorded the effective evolution 
of the state of an individual atom is reminiscent of one studied, e.g., in \cite{BCFFS}. 
In our derivation of these equations we make use of clear physical principles. Although our purpose, in this paper, is not to present many technical details, our analysis is based on mathematically rigorous reasoning.
Detailed proofs will appear elsewhere.\\

\noindent
Our paper is organized as follows.

In the next section, we introduce the physical systems and the models used to describe them. We verify that
these models fit perfectly into the formalism of the $ETH$ - Approach to QM. We derive a Lindblad equation
describing the effective time evolution of so-called \textit{``ensemble states,''} namely of averages of states over a large ensemble 
of identical, identically prepared systems. Subsequently, some properties of the relevant Lindblad equations are
summarized.

In Sect.~\ref{random walkers}, we sketch some key elements of the theory of diffusion for a gas of particles on a simple 
(hyper-) cubic lattice $\mathbb{Z}^{\nu}, \,\nu=1,2, \dots,$ and the associated theory of random walkers on 
$\mathbb{Z}^{\nu}$. The relation between the theory of diffusion, which is linear and deterministic, and the associated theory 
of random walkers, which is stochastic, turns out to be somewhat analogous to the relation between the quantum-mechanical 
time evolution of ensemble states, which is linear and deterministic, and the time evolution of states of individual systems, 
which is stochastic. The intention behind Sect.~\ref{random walkers} is to facilitate the readers' understanding of subsequent 
sections dealing with Quantum Mechanics.

In Sect.~\ref{Unraveling Lindblad} we derive the stochastic equations describing the effective
quantum-mechanical time evolution of states of individual atoms by ``unraveling'' the Lindblad 
equation describing the evolution of ensemble states presented in Sect.~\ref{Physical systems}. 
We assume that photons emitted by an atom are not recorded. Our analysis involves a novel form of 
perturbative analysis which we call ``infinitesimal perturbation theory.'' Sect.~\ref{Unraveling Lindblad} ends with a
sketch of how one may construct a measure on a space of atomic state trajectories exhibiting ``quantum jumps'' 
in analogy with the construction of a measure on the space of random walks parametrized  by time (similar to the 
\textit{Wiener measure} on Brownian paths).

Sect.~\ref{Q-Poisson} is dedicated to a derivation of the stochastic equations describing the evolution 
of states of atoms spontaneously emitting photons in the case where the emitted photons are recorded 
by photomultipliers. Our derivation is based on an ``unraveling'' of the Lindblad equation that takes into 
account the effect of detecting those photons and hence differs from the unraveling used in Sect.~\ref{Unraveling Lindblad};
(see also \cite{BCFFS}).

In Sect.~\ref{Example} we consider the special case of two-level atoms, the purpose being to present very explicit
results whose derivation is fairly straightforward.

Sect.~\ref{conclusions} contains some conclusions and an outlook. Among other things we draw attention to the fact
that, in the context of idealized models similar to the ones studied in this paper, the \mbox{$ETH$ -} Approach can be used
to describe measurements. Thus, this approach to QM eliminates various conundrums of text-book QM, such as
the so-called \textit{``measurement problem,''} or the \textit{``information paradox.''}

Various technical considerations are presented in three appendices.\\

\noindent
\textit{Acknowledgements.} 

We are grateful to \textit{Carlo Albert} and \textit{Henri Simon Zivi} for their stimulating interest in our efforts
and valuable comments. The senior author thanks \textit{Baptiste Schubnel} and \textit{Philippe Blanchard}
for past collaborations dedicated to developing the $ETH$ - Approach to Quantum Mechanics. ZG thanks the ETH for 
hospitality during his sabbatical stay in Zurich. AP acknowledges support through the MIUR Excellence Department Project 
awarded to the Department of Mathematics, University of Rome Tor Vergata, CUP E83C18000100006.

\section{Charged Matter Interacting with the Quantized Electromagnetic Field}\label{Physical systems}
The physical systems we propose to describe in the following consist of charged matter confined to a small region
of physical space interacting with the quantized electromagnetic field, possibly in the presence of photomultipliers 
that serve to detect photons spontaneously emitted by matter. We will focus our attention on idealized models 
of  static atoms (orbital motion neglected) interacting with the quantized radiation field, and we will study the limiting 
regime where the velocity of light, $c$, tends to $\infty$ (meaning that typical average velocities of atomic degrees of freedom 
are tiny as compared to $c$).
The state vectors of the atoms considered in the following are assumed to belong to a \textit{finite-dimensional} Hilbert space, 
$\frak{h}_A := \mathbb{C}^N$, with $N<\infty$. If decoupled from the radiation field the dynamics of an atom is generated 
by a selfadjoint Hamiltonian, $H_A= H_A^{*}$, acting on $\mathfrak{h}_A$, whose energy spectrum is given by $E_0<E_1<\dots<E_{N-1}$. 
The eigenvectors of $H_A$ form a complete orthonormal system, $\big\{\psi_0, \dots, \psi_{N-1}\big\}$, in $\mathfrak{h}_A$, with
\begin{equation}\label{atomic Ham}
H_A\psi_j = E_j \psi_j, \qquad \text{for  }\,\,j=0,1,\dots, N-1.
\end{equation}
The state space of the quantized electromagnetic/radiation field is the usual photonic Fock space 
denoted by $\mathcal{F}$. We introduce creation- and annihilation operators, $a^{*}$ and $a$, describing the emission 
or absorption of a photon. The Hamiltonian of the free radiation field is given by a selfadjoint operator, $H_f$, acting 
on $\mathcal{F}$; it is quadratic in $a^{*}$ and $a$. Fock space contains a unique (up to a phase) vector, called
\textit{vacuum} (vector), $\big|\emptyset \big>$, annihilated by all annihilation operators, $a$, and by the field Hamiltonian $H_f$.
Interactions between the (electrons in the shells of an) atom and the radiation field are described by an interaction Hamiltonian, 
$H_I=H_I^{*}$, acting on the tensor product space 
\begin{equation}\label{Hilbert space}
\mathfrak{H}:=\mathfrak{h}_A\otimes \mathcal{F}.
\end{equation}
The total Hamiltonian, $H$, is given by
\begin{equation}\label{Ham}
H:= H_0 + H_I, \quad \text{with}\quad H_0:= H_A \otimes \mathbf{1} + \mathbf{1}\otimes H_f \,.
\end{equation}
It is assumed to be a selfadjoint operator on $\mathfrak{H}$.
We suppose that the atom is located near the origin, $\mathbf{0}$, in physical space, and that interactions between
the atom and the radiation field (described by $H_I$) take place near $\mathbf{0}$. 

Photomultipliers are systems with infinitely many degrees of freedom. Their state space is an infinite-dimensional Hilbert space, denoted
$\mathcal{F}_{pm}$. They have the property that initial states in a large subspace of $\mathcal{F}_{pm}$ rapidly relax 
(at least \textit{locally}) towards a specific state $\big| s \big>$ corresponding to a ``silent,'' or ``sleeping,'' photomultiplier. 
For the purposes of this paper, a photomultiplier can be described in terms of some infinite subset of modes of the quantized 
electromagnetic field of low energy that are emitted when the photomultiplier is hit by an incident photon spontaneously emitted 
by an atom. More precisely, we may imagine that the photonic Fock space $\mathcal{F}$ is a tensor product space,
$$\mathcal{F}= \mathcal{F}_{pm} \otimes \mathcal{F}_A\,,$$
where $\mathcal{F}_A$ is the Fock space of states that can be occupied by photons of comparatively higher energy emitted
by an atom. The state $\big| s\big>$ can then be identified with the \textit{vacuum} in $\mathcal{F}_{pm}$. 
In the following, the dynamics of photomultipliers and their interactions with photons spontaneously emitted by an atom 
will not appear explicitly in our formulae; they do, however, play an important, albeit implicit, role, as we will see later.

\subsection{Models to be studied}
In this subsection we introduce some mathematical structure useful to render our analysis precise.\footnote{The reader may skip this subsection in
a first reading and proceed to Subsect.~2.2.} 
We first consider the free radiation field in the absence of atoms and photomultipliers.
Let $V^{+}_t$ denote the forward light cone in space-time with apex in the space-time point $(\mathbf{0}, t)$, where $t\in \mathbb{R}$ 
denotes time. By $\mathcal{A}_{\geq t}$ we denote the (von Neumann) algebra generated by all bounded functions of the 
electromagnetic field operators, $\big\{F_{\mu\nu}\big\}$, smeared out with real-valued test functions supported in $V^{+}_t$. 
It is well known that the free electromagnetic field obeys \textit{Huygens' Principle} in the form
\begin{equation}\label{Huygens}
(\mathcal{A}_{\geq t'})^{'} \cap \mathcal{A}_{\geq t} = \mathcal{A}_{[t,t']}\,, \quad \text{for }\,\, t'>t,
\end{equation}
where $\mathcal{A}_{[t,t']}$ is the infinite-dimensional  (von Neumann) algebra generated by all bounded functions of the 
electromagnetic field operators smeared out with real-valued test functions supported in the ``diamond,'' $D_{[t,t']}$, 
defined to be the intersection of the forward light cone $V^{+}_t$  with the backward light cone with apex in $(\mathbf{0}, t')$. 
The algebra $(\mathcal{A}_{\geq t'})^{'}$ is the so-called \textit{commutant} of $\mathcal{A}_{\geq t'}$, which is defined as follows: 
if $\mathcal{B}$ is an algebra of bounded operators acting on a Hilbert space 
$\mathcal{H}$ then its commutant, denoted by $\mathcal{B}'$, is the algebra of all bounded operators on $\mathcal{H}$ commuting 
with all operators in $\mathcal{B}$; (in Eq.~\eqref{Huygens}, $\mathcal{H}= \mathcal{F}$ and $\mathcal{B}= \mathcal{A}_{\geq t'}$). 
See \cite{Buchholz} for a general analysis of Huygens' Principle in quantum electrodynamics.

When interactions of the radiation field with an atom located near $\mathbf{0}$ are turned on Huygens' Principle, as formulated
in \eqref{Huygens},  does usually not hold, anymore, because one has to introduce an ultraviolet cutoff in the interaction Hamiltonian $H_I$ 
to tame ultraviolet (short-distance) singularities, and this invalidates \eqref{Huygens}.\footnote{This problem arises in the Pauli-Fierz
model of quantum electrodynamics with non-relativistic matter.} There are two possibilities 
to save Huygens' Principle: 
\begin{itemize}
\item{One may consider discretizing time; i.e., one replaces the time axis $\mathbb{R}$ by $\mathbb{Z}_{dt}$, where $dt>0$ 
is an elementary time step. This eliminates short distance singularities. It has been explored in some detail in our papers 
\cite{FP, FGP}.
}
\item{One may study the limiting regime where the velocity of light, $c$, tends to $\infty$, so that a light cone $V^{+}_t$ opens up to 
the half space $\big\{(\mathbf{x}, t')\,\big|\, \mathbf{x}\in \mathbb{E}^{3}, t'\geq t\big\}$ in (Newtonian) space-time. In this regime, one 
can let the time step $dt$ approach 0. The algebras $\mathcal{A}_{[t,t']}$ are then given by \textit{``time-slice algebras''}  for which 
a limiting form of Huygens' Principle holds. In this paper we are primarily concerned with the study of models in the limiting regime 
$c\rightarrow \infty$, which will now be described in more detail; (see also \cite{FGP}).}
\end{itemize}
In the limit where $c\rightarrow \infty$ the creation- and annihilation operators of the radiation field are given by
operators
\begin{align}\label{CCR}
\begin{split}
&a^{*}(t, \xi),\, a(t, \xi),\,\, t\in \mathbb{R},\, \xi \in \mathfrak{X}, \quad \text{ satisfying}\\
\big[a(t, \xi), a^{*}(t', \xi')\big] = &\delta(t-t') \,C(\xi, \xi')\,, \quad \big[a^{\#}(t, \xi), a^{\#}(t', \xi')\big]=0\,, \,\,\forall\,\, t, t', \xi, \xi'\,,
\end{split}
\end{align}
where $t\in \mathbb{R}$ is time, $\xi \in \mathfrak{X}$ denotes the state of a photon (i.e., a function of polarization and wave vector), 
$C$ is a bounded, non-negative quadratic form on the space $\mathfrak{X}$ of single-photon states, and $a^{\#} = a \text{ or } a^{*}$. 
We require the Heisenberg equations
\begin{equation}\label{Heisenberg}
e^{it(H_f/\hbar)} \,a^{\#}(t', \xi)\, e^{-it(H_f/\hbar)} = a^{\#}(t'+t, \xi)\,, 
\end{equation}
where $H_f$ is the Hamiltonian of the radiation field, and $\hbar$ is Planck's constant, which will henceforth be set to 1. 
Eqs. \eqref{CCR} and \eqref{Heisenberg} determine the form of $H_f$. It turns out that the spectrum of $H_f$ covers 
the entire real line, (a somewhat unphysical feature of the model). 
The \textit{vacuum} (vector), $\big| \emptyset\big>\in \mathcal{F}$ has the property that
\begin{equation}\label{vacuum}
a(t, \xi) \big|\emptyset \big> = 0, \,\,\, \forall\,\,\, t\in \mathbb{R},\, \forall\,\,\, \xi\in \mathfrak{X}\,.
\end{equation}
It follows from the commutation relations \eqref{CCR} that, given an arbitrary time $t\in \mathbb{R}$, Fock space factorizes as
\begin{align}\label{factorization}
\begin{split}
&\mathcal{F}= \mathcal{F}_{< t} \otimes \mathcal{F}_{\geq t}, \quad \text{with}\\
a^{\#}(t', \xi) = \,&a^{\#}(t', \xi)\big|_{\mathcal{F}_{< t}} \otimes \mathbf{1}\big|_{\mathcal{F}_{\geq t}}, \,\,\forall\,\, t' < t\, ,\\
a^{\#}(t', \xi) = \,&\mathbf{1}\big|_{\mathcal{F}_{< t}} \otimes a^{\#}(t', \xi)\big|_{\mathcal{F}_{\geq t}} , \,\,\forall \,\,t' \geq t\, .
\end{split}
\end{align}
By $\big|\emptyset, t\big>$ we denote the vector in $\mathcal{F}_{\geq t}$ with the property that
\begin{equation}\label{t-vacuum}
a(t', \xi) \big| \emptyset, t\big> = 0, \quad \forall\,\, t' \geq t, \, \forall\,\, \xi \in \mathfrak{X}\,.
\end{equation}
The space $\mathcal{F}_{[t, t')}, \text{ for }\, t'>t,$ is defined as the intersection of $\mathcal{F}_{\geq t}$ with $\mathcal{F}_{< t'}$.

Next, we describe the interaction Hamiltoian $H_I$. We assume that there are certain non-trivial transitions, $\psi_j \rightarrow \psi_i$, 
between internal states of the atom, accompanied by the absorption or emission of a photon in an appropriate photonic state 
$\xi_{ij} \in \mathfrak{X}$. We introduce operators, $D_{ij}$, on $\mathfrak{h}_A$, called transition amplitudes, with the properties that
\begin{equation}\label{transition amp}
\big< \psi_k, D_{ij} \psi_{\ell}\big> = \delta_{ki} \delta_{j\ell} \cdot d_{ij}\,, \qquad D_{ij}^{*}= D_{ji}\,,
\end{equation}
where the quantities $d_{ij}\in \mathbb{C}$ are non-zero for a certain set, $\mathcal{T}$, of pairs $(ij)$ corresponding to ``allowed''
atomic transitions. A simple (or simplistic) example of an interaction Hamiltonian $H_I$ is given by
\begin{equation}\label{interaction Ham}
H_I:= e \sum_{(i j)\in \mathcal{T}: j>i} \big[D_{ij}\otimes a^{*}(0, \xi_{ij})\, +\, D_{ji} \otimes a(0, \xi_{ij})\big]\,,
\end{equation}
where $e$ is the elementary electric charge. (The explicit form of the interaction Hamiltonian does not play a role in our theory.)
\vspace{0.15cm}\\
\textit{Simplifying assumptions:} 
\begin{itemize}
\item{For convenience, we will assume that the initial state of the radiation field prepared at some time $t=\underline{t}:=0$ 
is given by the vacuum state $\big| \emptyset, 0 \big> \in \mathcal{F}_{\geq 0}$. This simplifies the following analysis, 
but can be relaxed at the price of more complicated notations; see \cite{FP}.}
\item{Furthermore, we assume that the states 
$$a^{*}(\cdot, \xi_{ij})\big|\emptyset \big>\,\, \text{ and }\,\, a^{*}(\cdot, \xi_{k\ell})\big|\emptyset \big>$$ 
are orthogonal to one another whenever $i \not= k$, with $(ij), (k\ell) \text{ in } \mathcal{T}$, $ j>i$ and $ \ell>k$. 
Moreover, the states of the photomultiplier emerging when it is hit by photons in states $\xi_{ij}$ and $\xi_{k\ell}$ 
are assumed to be orthogonal to one another whenever $i \not= k$. These assumptions imply that the state of the radiation field 
after spontaneous emission of a photon is entangled with the state of the atom after emission.}
\end{itemize}
These assumptions are made for convenience; they can be relaxed at the price of complicating notations. 
They enable us to eliminate the radiation field and the photomultiplier from the analysis of the effective time 
evolution of the states of the atom, as we will see shortly.

In accordance with the $ETH$ - Approach to Quantum Mechanics (QM) described in \cite{FS, BFS, FP, FGP}, we introduce
algebras $\mathcal{E}_{\geq t}$ generated by all \textit{potential events/potentialities} (see \cite{FP, FGP}) that may happen at 
future times $\geq t, t\in \mathbb{R}$. We suppose that a system described by the model introduced above is prepared in some 
initial state at a time $t=\underline{t} = 0$. We set
\begin{equation}\label{initial alg}
\mathcal{E}_{\geq 0}:= B(\mathfrak{h}_A) \otimes B(\mathcal{F}_{\geq 0})\,,
\end{equation}
where $B(\mathcal{H})$ is the algebra of all bounded operators on a Hilbert space $\mathcal{H}$.
The algebra $\mathcal{E}_{\geq t}$ of \textit{potentialities} that may happen at future times $\geq t$ 
is then given by (see \cite{BFS, FP, FGP})
\begin{equation}\label{future alg}
\mathcal{E}_{\geq t}:= \big\{e^{itH} \, X \, e^{-itH}\,\big| \, X\in \mathcal{E}_{\geq 0}\big\} = e^{itH}\,\mathcal{E}_{\geq 0}\, e^{-itH}\,,
\end{equation}
where $t>0$ is arbitrary, and $H=H_0 + H_I$ is the Hamiltonian of the system, with $H_0$ defined in \eqref{Ham}.

Using \eqref{Heisenberg} we see that
\begin{align}\label{future alg-1}
\mathcal{E}_{\geq t}= U_t \big(B(\mathfrak{h}_A)\otimes B(\mathcal{F}_{\geq t})\big) U_t^{*}\,, \quad \text{ where }\,\, \,U_t := e^{itH}\cdot e^{-itH_f}\,.
\end{align}
By mimicking arguments presented in \cite{FP} for models with discrete time (i.e., $dt>0$) one shows that,
for an arbitrary $t$, the operator $U_t$ satisfies 
\begin{equation}\label{U}
U_t \in B(\mathfrak{h}_A)\otimes B(\mathcal{F}_{< t})\,.
\end{equation}
Actually, $U_t$ is unitary on $\mathfrak{h}_A\otimes \mathcal{F}_{< t}$, and hence, by \eqref{CCR}, it commutes with all operators in 
$\mathbf{1}\otimes B(\mathcal{F}_{\geq t})$.

The definition \eqref{future alg} of the algebras $\mathcal{E}_{\geq t}$ implies that
\begin{equation}\label{endo}
\mathcal{E}_{\geq (t'+t)} = e^{itH}\, \mathcal{E}_{\geq t'}\, e^{-itH}\,.
\end{equation}
Furthermore, Eqs.~\eqref{future alg-1}, \eqref{U} and  \eqref{factorization} imply that the family of algebras 
$\big\{\mathcal{E}_{\geq t}\big\}_{t\in \mathbb{R}}$ satisfies the \textit{``Principle of Diminishing Potentialities''} (PDP),
which says that, \textit{for arbitrary} $t'>t$,
\begin{align}\label{PDP}
\begin{split}
&\quad \mathcal{E}_{\geq t'}\subsetneq \mathcal{E}_{\geq t}, \quad \text{ and }\\
\big(\mathcal{E}_{\geq t'}\big)' &\cap \mathcal{E}_{\geq t} = U_{t'}\,\big(\mathbf{1}\otimes B(\mathcal{F}_{[t, t')}) \big) U_{t'}^{*}\,,
\end{split}
\end{align}
see \cite{FP}.
Note that the algebra on the right side of the equality sign in \eqref{PDP} is \textit{infinite-dimensional}.
The Principle of Diminishing Potentialities is a cornerstone of the $ETH$ - Approach to QM.\\

The way \textit{states} are defined in the $ETH$ - Approach is a key feature of this theory.\\
\textbf{Definition of States}: A \textit{state,} $\omega_t$, at time $t$ of the system described by the model introduced above 
(see \eqref{CCR} through \eqref{future alg}) is given by a \textit{normal state} (in the usual mathematical sense of this expression) 
on the algebra $\mathcal{E}_{\geq t}$ of potentialities in the future of time $t$, i.e., by a positive, normalized (weakly continuous) 
linear functional on $\mathcal{E}_{\geq t}$. 
We are interested in describing the dependence of $\omega_t$ on time $t$, given the time evolution
generated by the Hamiltonian $H$ and an initial state $\omega_0$ at time $t=\underline{t} := 0$. For this purpose we have to discriminate 
between \textit{ensemble states} and  \textit{states of individual systems}. Let $\mathfrak{E}_{\omega_0}$ denote an ensemble 
of identical systems, all described by the model introduced above, all of which are prepared in the same initial state $\omega_0$ 
at time $t=\underline{t}=0$. An ensemble state, $\omega_t$, at time $t$ is an average over \textit{all possible histories} of states at times $t'\in [0,t]$ 
of \textit{individual systems} in $\mathfrak{E}_{\omega_0}$ prepared in the same initial state $\omega_0$. In the 
$ETH$ - Approach, the ensemble state $\omega_t$ at time $t>0$ is given by restricting the initial state $\omega_0$ to the subalgebra 
$\mathcal{E}_{\geq t}$ of the algebra $\mathcal{E}_{\geq 0}$; i.e., 
\begin{equation}\label{ensemble state}
\omega_t:= \omega_0 \big|_{\mathcal{E}_{\geq t}}\,.
\end{equation}
Note that, because of the Principle of Diminishing Potentialities, the state $\omega_t$ will usually be a \textit{mixed} state on 
$\mathcal{E}_{\geq t}\, (\subsetneq \mathcal{E}_{\geq 0})$ \textit{even if} $\omega_0$ is a \textit{pure} state on $\mathcal{E}_{\geq 0}$ 
(a manifestation of entanglement !).

It follows from the Heisenberg equations for the operators in the algebras $\mathcal{E}_{\geq t}$ (see \eqref{Heisenberg} and 
\eqref{future alg} through \eqref{U}) that the time evolution of ensemble states is determined by a \textbf{linear deterministic equation}, 
which, for the model described above, turns out to be a \textit{Lindblad equation} \cite{Lindblad}. 
In contrast, the time evolution of states of \textit{individual systems} in $\mathfrak{E}_{\omega_0}$ is given 
by a \textbf{non-linear stochastic equation}, whose form depends on whether the photomultiplier is present and turned on or
is absent. This equation can be derived from the evolution equation for ensemble states by systematically applying the 
\textit{rules of the $ETH$ - Approach,} as formulated in \cite{BFS, FP, FGP}, viz.~the \textit{Principle of Diminishing Potentialities} 
and the so-called \textit{State-Reduction Postulate} (Axiom CP of \cite{FGP}). We will not present all the details of this derivation
in the present paper. For models with discrete time, i.e., with a positive time step $dt$, this has been done in \cite{FP}; the limit as $dt\searrow 0$
is analyzed in \cite{FGP-2}. In Sects.~\ref{Unraveling Lindblad} and \ref{Q-Poisson}, some of the key ideas of this derivation are outlined.

\subsection{Effective evolution equation for ``ensemble states'' of atoms}\label{eff evol}

To keep our analysis simple, we will henceforth assume that the initial state of the radiation field is the vacuum state 
given by the rank-1 projection $\big| \emptyset \big>\big< \emptyset \big|$, where $\big| \emptyset \big>$ is the 
vacuum vector in Fock space $\mathcal{F}$.
The initial state of the atom is given by a density matrix on $\mathfrak{h}_A$, i.e., by a positive $N\times N$-matrix, $\Omega_0$, with
$\text{Tr}(\Omega_0)=1$. Hence the initial state $\omega_0$ of the systems under study is given by
\begin{equation}\label{initial state}
\omega_0 = \Omega_0 \otimes \big| \emptyset \big>\big< \emptyset \big|\,.
\end{equation}
This is a normal state on the algebra $\mathcal{E}_{\geq 0}$ defined in \eqref{initial alg}. Using \eqref{future alg-1}, \eqref{U} 
and \eqref{factorization}, one may then show that, up to unitary conjugation with the operator $U_t$, the ensemble state 
$\omega_t$ on the algebra $\mathcal{E}_{\geq t}$ corresponding to the initial state $\omega_0$ is given by
\begin{equation}\label{states}
\omega_t = \Omega_t \otimes \big| \emptyset, t \big> \, \big<  \emptyset, t \big|\,,
\end{equation}
where, we recall, $\big| \emptyset, t\big>$ is the vacuum vector in $\mathcal{F}_{\geq t}$.
Thus, for an initial state $\omega_0$ as in \eqref{initial state}, the time evolution of ensemble states
is entirely determined by the evolution, $\Omega_0 \rightarrow \Omega_t$, of the \textit{density matrix} 
on the Hilbert space $\mathfrak{h}_A$ of atomic degrees of freedom. 

Assuming that $H_f$ is as in \eqref{Heisenberg}, that the interaction Hamiltonian $H_I$ is given by \eqref{interaction Ham}, one shows
that the time dependence of $\Omega_t$ is given by a \textit{Lindblad equation} \cite{Lindblad}
\begin{align}\label{Lindblad}
\begin{split}
&\dot{\Omega}_t = \mathfrak{L}_{\alpha}[\Omega_t]\,, \quad \text{ where }\,\,\dot{\Omega}_t := \frac{\partial \Omega_t}{\partial t},\, \text{ and}\\
\mathfrak{L}_{\alpha}[\Omega]:= -&i \big[H_A, \Omega \big] + \alpha \sum_{(ij)\in \mathcal{T}:j>i} \Big[D_{ij} \Omega D_{ij}^{*} - \frac{1}{2} 
\big\{\Omega, D_{ij}^{*}\,D_{ij}\big\} \Big]\,,
\end{split}
\end{align}
where $\Omega$ is an arbitrary density matrix on $\mathfrak{h}_A$, $\alpha \propto e^{2}$ is a positive constant, 
$\big[A,B\big]:= A\cdot B - B\cdot A$ is the commutator, and $\big\{A,B\big\}:= A\cdot B + B\cdot A$ is the anti-commutator 
of two operators $A$ and $B$.

\textit{Remark:} The methods developed in this paper can actually be applied to very general evolution equations for ensemble states 
of the form of \eqref{Lindblad}, including ones with memory terms, with $\mathfrak{L}_{\alpha}$ replaced by
$$ \mathfrak{L} \equiv \mathfrak{L}[\cdot, t], \quad \text{ with } \,\,\, \text{Tr}\big(\mathfrak{L}[\cdot, t]\big) =0\,,$$ 
where $\mathfrak{L}[\cdot, t]$ is a general linear operator acting on a \textit{space of trajectories of states,} 
$\big\{\Omega_{t'}\,\big|\,  \underline{t}\leq t'\leq t\big\}$, and the system is prepared in an initial state $\Omega_0$ 
at time $t=\underline{t}=0$.

\subsection{Some properties of solutions of the Lindblad equation}\label{return to gs}

It is of interest to discuss some properties of solutions of the Lindblad equation \eqref{Lindblad} under suitable 
conditions on the generator $\mathfrak{L}_{\alpha}$ relevant for the results described in subsequent sections. 

In Subsect.~\ref{eff evol} we have assumed that the radiation field is prepared in its vacuum state $\big|\emptyset \big>\big< \emptyset \big|$. 
In this case the operator $D_{ij}$ appearing on the right side of \eqref{Lindblad} describes a decay of the atom from a state 
$\psi_j$ to a state $\psi_i$ of lower energy (i.e., $j>i$), for each pair $(ij)$. We imagine that, 
for every \mbox{$j=1, \dots, N-1,$} where $N=\text{dim}\mathfrak{h}_A$, there exists at least one sequence, 
$(0, i_1), (i_1, i_2), \dots, (i_k, j),$ with\,$ j> i_k> \dots > i_1> 0,$ for some $k=1,\dots, N-2$, of allowed atomic transitions, 
(i.e., pairs belonging to the set $\mathcal{T}$) connecting $j$ to $0$, meaning that an atom prepared in a state $\psi_j$ 
may end up in the groundstate $\psi_0$ after a cascade of at most $N-1$ decay processes; in particular, the operators 
$D_{0i_1}, D_{i_1 i_2}, \dots, D_{i_k j}$ must all be non-zero. Under these hypotheses, every solution $\Omega_t, t\in \mathbb{R},$ 
of the Lindblad equation \eqref{Lindblad} has the property that
\begin{equation}\label{R to gs}
\Omega_t \rightarrow \big|\psi_0\big>\big< \psi_0\big|, \quad \,\,\text{ as }\,\, t\rightarrow \infty\,.
\end{equation}
The convergence in \eqref{R to gs} is called \textit{``relaxation to the groundstate''}, a well known phenomenon
in the quantum theory of atoms coupled to the radiation field, with the latter prepared in the vacuum state; 
(see \cite{FGS, DeR-K} for mathematical results on relaxation to the groundstate in models with a finite 
velocity of light). The convergence in \eqref{R to gs} can be established by expanding
$$\Omega_t= \text{exp}\big(t\, \mathfrak{L}_{\alpha}\big)\big[\Omega_0\big], \quad \Omega_{t=0}=\Omega_0,$$
in a Dyson series in powers of the term $\mathfrak{L}_{\alpha}''$, where
\begin{align}\label{21'}
\begin{split}
\mathfrak{L}_{\alpha} &= \mathfrak{L}_{\alpha}' + \mathfrak{L}_{\alpha}'', \quad \text{ with }\\
\mathfrak{L}_{\alpha}'[\Omega]&:= -i \big[H_A, \Omega \big] - \alpha \sum_{(ij)\in \mathcal{T}:j>i} \frac{1}{2}\big\{\Omega, D_{ij}^{*}\,D_{ij}\big\}\,,\\
\mathfrak{L}_{\alpha}'' &:= \alpha \sum_{(ij)\in \mathcal{T}:j>i} D_{ij} \Omega D_{ij}^{*}\,,
\end{split}
\end{align} 
and then using the conditions on atomic transitions formulated above.

We define the (von Neumann) entropy of a state $\Omega$ by
\begin{equation}\label{vN entropy}
S(\Omega):= - \text{Tr}\big( \Omega\cdot \ln\,\Omega\big)\,,
\end{equation}
where ``$\ln$'' denotes the natural logarithm. We note that $S(\Omega) \geq 0$.
The state $\Omega$ is \textit{pure}, i.e., given by a rank-1 orthogonal projection on $\mathfrak{h}_A$, iff $S(\Omega)=0$;
$\Omega$ is \textit{mixed} iff $S(\Omega)>0$.
We observe that \textit{even} if the initial state $\Omega_0$ is \textit{pure} the states $\Omega_t, t>0,$ solving 
\eqref{Lindblad}, with $\Omega_{t=0}=\Omega_0$, are \textit{mixed}, i.e., $S(\Omega_t)>0$. However, 
property \eqref{R to gs} implies that 
$$S\big(\Omega_t\big) \rightarrow 0, \qquad \text{as }\,\, t\rightarrow \infty.$$
Thus, if the initial state $\Omega_0$ is pure then the entropy $S(\Omega_t)$ must increase in $t$, for $t$ small enough,
but approaches $0$, as $t\rightarrow \infty$.\footnote{I.e., $S\big(\Omega_t\big)$ follows a ``Page Curve'' \cite{Page}, 
known in atomic physics apparently before it was noted in black-hole physics.}

We emphasize that the methods developed in this paper can also be used to study the time evolution of atoms interacting
with the radiation field when the latter is prepared in states containing photons at all times, and/or when the atom exhibits Rabi 
oscillations, or when the atom interacts with measurement devices. We will however not present such generalizations in the
present paper; (but see \cite{FP}, where they are studied for models with discrete time, $dt>0$).

\subsection{Goal of analysis}\label{Goal}
The goal of this paper is to apply the state-reduction postulate of the $ETH$ - Approach (Axiom CP of \cite{FGP}) to \textit{``unravel''} 
the Lindblad evolution of Eq.~\eqref{Lindblad}, so as to obtain a \textit{stochastic differential equation} describing the time evolution of states of 
\textit{individual} atoms prepared in an initial state \mbox{$\Omega_{t=\underline{t}}=\Omega_0$.}
The ``unraveling'' of the linear deterministic time evolution of \textit{ensemble states} will turn out to depend on whether photons 
emitted by the atom are recorded by a photomultiplier, or not. General results will be presented in Sections \ref{Unraveling Lindblad} 
and \ref{Q-Poisson}, the example of two-level atoms will be treated very explicitly in Sect.~\ref{Example}. (Readers not interested
in the general theory might want to directly proceed to Sect.~\ref{Example}.) The formalism developed in this paper enables us to describe
processes such as the fluorescence of atoms, more generally the decay of unstable states of matter, measurements of physical 
quantities represented by self-adjoint operators (see \cite{FP}), etc.

In order to prepare the ground for our analysis in Sects.~\ref{Unraveling Lindblad} - \ref{Example}, we consider the diffusion equation for the density of particles hopping on 
a simple (hyper-) cubic lattice, $\mathbb{Z}^{\nu}$, whose ``unraveling'' is given by the theory of simple random walks 
on $\mathbb{Z}^{\nu}$ parametrized by time, i.e., by a \textit{Poisson jump process} for random walkers hopping on 
$\mathbb{Z}^{\nu}$. This is the contents of the next section. The diffusion equation is the analogue of the Lindblad equation
\eqref{Lindblad}, and the Poisson jump process of a random walker on $\mathbb{Z}^{\nu}$ is the analogue of the stochastic 
process describing the quantum-mechanical evolution of states of individual systems described by the models introduced 
above.

\section{Diffusion and the Theory of Random Walkers on $\mathbb{Z}^{\nu}$}\label{random walkers}
In this section we introduce, in a simple and familiar context, some of the ideas that will become important in later
sections of this paper. 
 
We consider a system consisting of a large ensemble, $\mathfrak{E}$, of identical random walkers on the 
simple (hyper-) cubic lattice $\mathbb{Z}^{\nu}$ of dimension $\nu=1,2,3,\dots$. 
An \textit{ensemble state} of this system is given by a density $\rho$ on $\mathbb{Z}^{\nu}$, with $\rho(x)\geq 0, \,\,\forall\,\,x\in \mathbb{Z}^{\nu}$, which we normalize such that
$$\sum_{x\in \mathbb{Z}^{\nu}} \rho(x)=1\,.$$
The density $\rho=\rho_t$ may depend on time $t$; we assume that its time-dependence is governed by the diffusion equation
\begin{equation}\label{diffusion}
\dot{\rho}_t (x) = D \big(\Delta \rho_t\big)(x) = D \big[\sum_{y: |y-x|=1} \rho_t (y)\big] - 2\nu D\, \rho_t(x)\,,
\end{equation}
where $\dot{\rho}_t := \partial \rho_t/ \partial t$, $\Delta$ is the discrete Laplacian, $D$ is the diffusion constant, 
and the sum on the right side extends over all sites $y$ that are nearest neighbors of the site $x$, which we indicate by writing $|y-x|=1$.

One easily sees that if $\rho_t$ satisfies the diffusion equation \eqref{diffusion} and $\rho_0$ is a non-negative function on $\mathbb{Z}^{\nu}$ 
then $\rho_t$ is a non-negative function, for all times $t\geq 0$, (this is a consequence of the fact that the heat kernel is positivity preserving); 
moreover, $\sum_{x\in \mathbb{Z}^{\nu}} \rho_t(x)$ is \textit{independent} of time $t$, hence if $\sum \rho_0(x)=1$ then 
$\sum \rho_t(x)=1$, for all $t\geq 0$. The \textit{``entropy''}, $S$, of an ensemble state $\rho$ is defined by
\begin{equation}\label{entropy}
S(\rho):= -\sum_{x\in \mathbb{Z}^{\nu}} \rho(x)\cdot \ln \rho(x)\,.
\end{equation}
Using ``summation by parts,'' the mean-value theorem and the fact that $\rho_t(x)\geq 0, \forall\, x\in \mathbb{Z}^{\nu}$, 
one easily shows that if $\rho_t$ satisfies the diffusion equation then $S(\rho_t)$ is a \textit{monotone-increasing} function of time $t$,
which confirms the dissipative nature of diffusion.

We would like to understand what kind of stochastic motion of a single random walker $\omega \in \mathfrak{E}$ implies
that the time-dependence of an ensemble state $\rho_t$ is determined by the diffusion equation \eqref{diffusion}. For this purpose, we
have to \textit{``unravel''} \eqref{diffusion}. It helps intuition to re-write this equation in the form
\begin{equation}\label{diffusion-1}
\rho_{t+dt}(x) = \rho_{t}(x) + D\big(\Delta \rho_t\big) \cdot dt + \mathcal{O}(dt^2)\,, 
\end{equation}
where $dt$ is assumed to approach 0. 

In order to ``unravel'' \eqref{diffusion-1}, we first have to clarify the \textit{``ontology''} of random walkers.
The state of a single random walker $\omega\in \mathfrak{E}$ at an arbitrary time $t$ is a site \mbox{$x_{\omega}(t)\in \mathbb{Z}^{\nu}$.} 
It can be represented by the density $\rho_{ t} (x; \omega):= \delta_{x_{\omega}(t)}(x), \, x\in \mathbb{Z}^{\nu}$, where 
\mbox{$\delta_{y}(x) = 1$} \text{ if } $x=y$, and $\delta_{y}(x)=0 \text{ if } x\not=y$.  Assuming that $dt$ is very small, then $\omega$ may remain at the site $x_{\omega}(t)$ during the time interval $[t,t+dt)$, or it may jump to a nearest-neighbor site $y$, 
with $\big|y - x_{\omega}(t)\big| =1$; i.e., during the time interval $[t,t+dt)$,  its state may jump from $\delta_{x_{\omega}(t)}$ 
to $\delta_{y}$, with $\big|y - x_{\omega}(t)\big| =1$. The diffusion equation \eqref{diffusion-1} enables us to determine the probabilities 
for $\omega$ to remain in the state $\delta_{x_{\omega}(t)}$ or to jump to a state $\delta_{y}$, with $\big|y - x_{\omega}(t)\big| =1$,
during the time interval $[t,t+dt)$. We write \eqref{diffusion-1} as
\begin{equation}\label{mixture}
\rho_{t+dt}(x) = \big[1- 2D\,\nu\, dt\big] \rho_t(x) + \sum_{y: |y-x|=1} \big[D\, dt\big] \rho_{t}(y) +\mathcal{O}(dt^2)\,,
\end{equation}
where $\rho_t(x)=\rho_t(x; \omega)=\delta_{x_{\omega}(t)}(x)$.
Apparently, $\rho_{t+dt}$ is a convex superposition of ``pure states'' corresponding to sites $x_{\omega}(t)$ and 
\mbox{$y=x_{\omega}(t) + \delta$,} where $\delta$ is a lattice unit vector (i.e., $|y-x_{\omega}(t)|=|\delta|=1$). 
The coefficients appearing in this superposition can be interpreted as \textit{probabilities.} 
Thus, assuming that $dt$ is very small, the state $\rho_{t+dt}(\cdot; \omega)$ of the random walker $\omega$ at time $t+dt$ is given by
\begin{equation}\label{Purification}
\rho_{t+dt}(\cdot;\omega)=\begin{cases}
\delta_{x_{\omega}(t)}, & \text{ with probability }\,\, p_{nj}[t, t+dt]:= 1- 2\nu D\, dt\,,\\
\delta_{x_{\omega}(t)+ \delta}, & \text{ with probability }\,\, p^{\delta}[t, t+dt]:= D\, dt \,\,(\text{independent of }\, \delta)\,,
\end{cases}
\end{equation}
where the subscript ``$nj$'' stands for ``no jump'' and the superscript ``$\delta$'' stands for a jump of the random walker 
by the lattice unit vector $\delta$. Note that
$$p_{nj}[t, t+dt] + \sum_{\delta} p^{\delta}[t, t+dt] = 1\,,$$
as expected. Suppose that a random walker $\omega$ starts at some site $\omega(0)\in \mathbb{Z}^{\nu}$ at time $t=\underline{t}:=0$ and
makes $n=\ell(\omega)$ nearest-neighbor jumps at times $t\in [t_k, t_k + dt_k], k=1, \dots, n,$ along a simple random walk also 
denoted by $\omega$ 
$$\omega:=\big\{\omega(0), \dots, \omega(n)\,\big|\, \omega(k)\in \mathbb{Z}^{\nu}, |\omega(k+1)-\omega(k)|=1, k=0, 1, \dots \big\}$$ 
of length $n=\ell(\omega)$, until it stops at some time $t=\overline{t}$. One would like to predict the probability of 
encountering such a random walker, given the probabilistic law in Eq.~\eqref{Purification}. To determine this probability we first have
to calculate the probability that $\omega$ remains at the site $\omega(k-1)\in \mathbb{Z}^{\nu}$ during the time interval $[t_{k-1}, t_k)$, 
which we denote by $p_{nj}[t_{k-1}, t_k]$. The result is obtained by noticing that, for $t'<t''$,
$$\ln \, p_{nj}[t', t''] = \int_{t'}^{t''} \ln\, p_{nj}[t, t+dt] \overset{\eqref{Purification}}{=} - 2\nu D \int_{t'}^{t''} dt = 
-2\nu D\big(t'' - t'\big)\,,$$
hence
\begin{equation}\label{no jump1}
p_{nj}[t', t'']= e^{- 2\nu D\,(t''-t')}\,.
\end{equation}
When combined with expression \eqref{Purification} for $p^{\delta}$ this equation yields the following formula for the
probability of encountering a random walker $\omega \in \mathfrak{E}$ that makes nearest-neighbor jumps at (random) times
$t_1, \dots, t_n$, with $\underline{t}:= 0<t_1, \dots, t_n< \overline{t}$, 
along a simple random walk $\omega$ in $\mathbb{Z}^{\nu}$:
\begin{align}\label{proba}
\begin{split}
W_{\omega}[t_1, &\dots, t_n; 0, \overline{t}\,] \prod_{k=1}^{n} dt_k=\\
& =\Big\{\prod_{k=1}^{n} p_{nj}[t_{k-1}, t_k] p^{\omega(k)- \omega(k-1)}[t_k, t_k + dt_k]\Big\} p_{nj}[t_n, \overline{t}]\\
& \overset{\eqref{no jump1}}{=} D^{n}\,e^{-2{\nu}D\, \overline{t}}\, dt_1 \dots dt_n\,.
\end{split}
\end{align}
Next, we outline some consequences of this law.
\begin{enumerate}
\item[(i)]{ Let $\Delta_{n}[0, \overline{t}]$ be the simplex $\big\{t_1, \dots, t_n\,\big| \,0< t_1 <\dots< t_n <\overline{t} \,\big\}$. Summing over all 
random walkers $\omega \in \mathfrak{E}$ starting at a site $y$ and ending at a site $x$, one finds that
\begin{equation}\label{heat kernel}
\sum_{n=0}^{\infty} \sum_{\omega: y\rightarrow x, \ell(\omega)=n} \int_{\Delta_n[0,\overline{t}]} W_{\omega}[t_1, \dots, t_n; 0,\overline{t}\,] 
\prod_{k=1}^{n} dt_k\, =\, \big(e^{\overline{t}D\Delta}\big)_{xy}\,,
\end{equation}
where $\big(e^{\overline{t}\Delta}\big)_{xy}$ is the ``heat kernel,'' i.e., the solution at time $t=\overline{t}$ of the diffusion equation 
\eqref{diffusion} with initial condition $\rho_0 =\delta_{y}$. Thus, the evolution of ensemble states is indeed given by \eqref{diffusion}.}
\item[(ii)]{In order to find out how far, in average, a random walker gets in a time $t$, we calculate his/her mean-square displacement.
\begin{align*}
\mathbb{E}_{\omega\in\mathfrak{E}}\big[x_{\omega}(t)&-\,x_{\omega}(0)\big]^{2}=\\
&= \sum_{n} \underbrace{\mathbb{E}\big[\omega(n)-\omega(0)\big]^{2}}_{=n} \frac{(2\nu D\cdot t)^{n}}{n!} \,e^{-2\nu D\cdot t}\\
& = 2\nu D\cdot t \\
& = \sum_{x\in \mathbb{Z}^{\nu}} \big(e^{tD\,\Delta}\big)_{xy} \big(x-y\big)^{2}\,, \qquad \text{by\, Eq.~\eqref{heat kernel}}.
\end{align*}
This result confirms the diffusive nature of the motion of random walkers.}
\end{enumerate}

The location $x_{\omega}(t)$ of a random walker $\omega\in \mathfrak{E}$ at an arbitrary time $t$ is called an \textit{``event''} (E),
the ensemble, $\mathfrak{E}_{x_0}$, of all random walkers starting to move at an arbitrary but fixed site $x_0$ of the lattice 
at time $t=0$ is called a \textit{``tree''} (T) (of random walkers rooted in $x_0$), and the trajectory 
$\big\{x_{\omega}(t)\,\big|\, 0\leq t \leq \overline{t}\,\big\}$ of a random walker $\omega \in \mathfrak{E}$ (i.e., a particular 
branch of the tree) is called a \textit{``history''} (H). This motivates the language, in particular the name 
``$ETH$ - Approach,'' used in \cite{BFS, FGP, FP} and in the following sections.

\section{Stochastic Quantum Evolution \textit{Without} Photon Detection}\label{Unraveling Lindblad}
\textit{``Several recent studies have shown that the time evolution of an atom submitted to coherent laser fields and to
dissipative processes, such as spontaneous emission of photons [...], can be considered to consist of a sequence of 
coherent evolution periods separated by quantum jumps occurring at random times.''} -- C.~Cohen-Tannoudji et al., in \cite{C-T-Zambon}.
 
In this section, we present results, derived from basic principles, that confirm this claim. We study the time evolution of 
states of an \textit{individual} physical system, $S$, consisting of an atom interacting with the quantized electromagnetic 
field in the limiting regime where the velocity of light $c\rightarrow \infty$; $S$ belongs to an ensemble, $\mathfrak{E}$, of 
identical systems all described by the model introduced in Sect.~2.1. We suppose that there are \textit{no} 
photomultipliers present recording photons emitted by the atom. (The dynamics of such systems when 
emitted photons are detected by photomultipliers is treated in the next section.)

We assume that the initial state of the radiation field is the vacuum state $\big|\emptyset\big>\big< \emptyset\big|$; 
(more general initial states can be treated, too, but render the analysis more complicated, see \cite{FP}). 
As argued in Sect.~2.2, the state of the radiation field at an arbitrary \textit{future} time $t$ is then given by the vacuum state 
$\big| \emptyset,t\big> \big<\emptyset,t\big|$, with $\big| \emptyset,t\big>\in \mathcal{F}_{\geq t}$, which is a pure state on 
the algebra $B(\mathcal{F}_{\geq t})$. For this reason, it suffices to consider the effective evolution of the state of the atom, 
which is given by a density matrix, $\Omega$, on the atomic Hilbert space $\mathfrak{h}_A$. The time dependence of an 
``ensemble state,'' $\Omega_t$, i.e., of an average of states of individual atoms over the ensemble $\mathfrak{E}$, is then 
governed by the Lindblad equation \eqref{Lindblad}, which we now re-write as follows:
\begin{align}\label{Lindblad-1}
\begin{split}
&\Omega_{t+dt}= \Omega_t + \mathfrak{L}_{\alpha}\big[\Omega_t\big]\,dt + \mathcal{O}(dt^2), \qquad \text{with }\\
\mathfrak{L}_{\alpha}\big[\Omega\big] &:= -i\big[H_A, \Omega\big] + \alpha \sum_{(ij)\in \mathcal{T}} \Big[D_{ij} \Omega D_{ij}^{*} - 
\frac{1}{2}\big\{\Omega, D_{ij}^{*} D_{ij}\big\}\Big]\,, 
\end{split}
\end{align}
where the operators (transition amplitudes) $D_{ij}$ have been defined in Eq.~\eqref{transition amp}. For the systems described 
by this model, the state-reduction postulate (Axiom CP in \cite{FGP}) of the \mbox{$ETH$ -} Approach to quantum 
mechanics implies that the state of an \textit{individual} atom is \textit{pure}, i.e., given by a rank-1 orthogonal projection, 
$\Pi_t = \Pi_{t}^{*}= \Pi_{t}^{2},$ on $\mathfrak{h}_A$, at \textit{all} times $t$.
Assuming that an atom in $\mathfrak{E}$ is prepared at time $t$ in a state given by a rank-1 orthogonal projection $\Pi_t$, the ensemble
state, $\Omega_{t+dt}$, at time $t+dt$, with $dt>0$ very small, (i.e., the state averaged over the ensemble $\mathfrak{E}_{\Pi_t}$ of systems 
all identical to $S$ and prepared in the \textit{same} pure state $\Pi_t$ at time $t$) is then given by
\begin{equation}\label{mixing states}
\Omega_{t+dt}= \Pi_t + \mathfrak{L}_{\alpha}\big[\Pi_t\big]\,dt + \mathcal{O}(dt^{2})\,.
\end{equation}
Here $\Omega_{t+dt}$ is a density matrix, i.e., a positive operator on $\mathfrak{h}_A$ with $\text{Tr}\big(\Omega_{t+dt}\big)=1$. 
According to the spectral theorem, it can be decomposed into a convex combination of pure states
\begin{equation}\label{convex combi}
\Omega_{t+dt} = p_{nj}[t, t+dt]\, \Pi_{t+dt}^{0} + \sum_{\delta=1}^{N-1} p^{\delta}[t, t+dt]\, \Pi_{t+dt}^{\delta}\,,
\end{equation}
where $\Pi_{t+dt}^{0}, \Pi_{t+dt}^{1}, \dots, \Pi_{t+dt}^{N-1}$ are mutually disjoint\footnote{two orthogonal projections
$\Pi_1$ and $\Pi_2$ are disjoint iff $\Pi_1\cdot \Pi_2 =0$.} orthogonal projections of rank 1, with
$$\sum_{\delta=0}^{N-1} \Pi_{t+dt}^{\delta} = \mathbf{1}\big|_{\mathfrak{h}_A},$$
and, assuming that $dt$ is small enough,
\begin{align}\label{probabilities}
\begin{split}
&p_{nj}[t, t+dt] \equiv p^{0}[t, t+dt] = 1- \mathcal{O}(dt) >0\,,\\
p_{nj}>&p^1\geq \dots \geq p^{N-1}\geq 0, \quad \text{with }\,\, p^{\delta} = \mathcal{O}(dt), \,\,\forall\,\, \delta \geq 1\,,\\
&\qquad p_{nj}[t, t+dt] + \sum_{\delta=1}^{N-1} p^{\delta}[t, t+dt] = 1\,,
\end{split}
\end{align}
where ``$nj$'' stands for ``no jump''. (For simplicity, we will ignore complications caused by degeneracies, $p^{\delta}= p^{\delta+1}$, 
for some $\delta\geq 1$, which are non-generic.) According to the $ETH$ - Approach, \textit{one} of the projections 
$\big\{ \Pi_{t+dt}^{\delta}\,\big|\, \delta = 0, \dots, N-1\big\}$, \textit{randomly chosen}, is the state of an \textit{individual} atom 
at time $t+dt$. (This is what, in \cite{BFS, FP, FGP}, is called an \textit{``actual event''} or \textit{``actuality''}.) 
The \textbf{``ontology''} proposed here is reminiscent of the one adopted in the theory of random walkers 
sketched in Sect.~\ref{random walkers}.
In the context of the $ETH$ - Approach, \textit{Born's Rule} says that the probability, or frequency of encountering $\Pi_{t+dt}^{\delta},$
for some $\delta = 0, \dots, N-1,$ is given by $p^{\delta}[t, t+dt]$, where $p^{0}[t, t+dt] = p_{nj}[t, t+dt]$.
These rules are analogous to the ones used in Sect.~\ref{random walkers}, Eqs.~\eqref{mixture} and \eqref{Purification}, 
to determine the motion of a random walker, and this has been our motivation to discuss diffusion and the theory of random walkers 
as a ``warm-up exercise'' for the analysis presented in the following.

In order to find explicit expressions for the objects appearing on the right side of \eqref{convex combi}, we make use of \textit{``infinitesimal
perturbation theory''} (see Appendix A), which we briefly sketch here.

\subsection{``Unraveling'' the Lindblad equation with the help of infinitesimal perturbation theory}\label{IAPT} 
Let $H_0$ be a symmetric linear operator on $\mathbb{C}^{N}$ (i.e.,
a symmetric $N\times N$ matrix) with a simple eigenvalue $E_0$ separated from the rest of its spectrum by a strictly positive gap. Let
$\Pi_0$ denote the orthogonal projection onto the eigenvector of $H_0$ corresponding to the eigenvalue $E_0$. Let $V$ be a linear
operator on $\mathbb{C}^{N}$, and consider the perturbed operator 
$$H_{\varepsilon}:= H_0 + \varepsilon V, \quad \text{ with }\,\, \varepsilon \ll 1\,.$$
We would like to have explicit expressions for the perturbed eigenvalue $E_{\varepsilon}$ of $H_{\varepsilon}$ growing out of $E_0$
and for the eigenprojection, $\Pi_{\varepsilon}$, onto the eigenvector of $H_{\varepsilon}$ corresponding to $E_{\varepsilon}$.
For this purpose, we define an operator $\mathcal{S}$ by
\begin{equation}\label{generator}
\mathcal{S}:= \big(H_0 - E_0\big)^{-1} \Pi_{0}^{\perp}\,V\, \Pi_0 - \Pi_0 \, V\, \Pi_{0}^{\perp} \big(H_0 - E_0\big)^{-1}\,,
\end{equation}
where $\Pi_0^{\perp} := \mathbf{1}-\Pi_0$. It is easy to verify that, up to corrections of order $\mathcal{O}(\varepsilon^{2})$, 
the operator $e^{\varepsilon \mathcal{S}}\, H_{\varepsilon} \, e^{-\varepsilon \mathcal{S}}$ is block-diagonal with respect to 
the projection $\Pi_0$ and its complement, and that the following formulae hold true. \vspace{0.15cm}\\
\textit{Formulae for perturbed eigenvalues and eigenstates:}
\begin{align}\label{PT}
\begin{split}
E_{\varepsilon} =& E_0 + \varepsilon\, \text{Tr}( \Pi_0\, V) + \mathcal{O}(\varepsilon^{2})\\
\Pi_{\varepsilon} =& \Pi_0 - \varepsilon \big[\mathcal{S}, \Pi_0\big] + \mathcal{O}(\varepsilon^{2})\,.
\end{split}
\end{align}
\textit{Assuming that $H_{\varepsilon}= H_{\varepsilon}^{*}$ is a selfadjoint operator then its eigenvalue spectrum, except for $E_{\varepsilon}$, 
and the corresponding spectral projections can be found by diagonalizing the operator $\Pi_{0}^{\perp}\, H_{\varepsilon}\,\Pi_{0}^{\perp}$, 
up to corrections of $\mathcal{O}(\varepsilon^{2})$.}\vspace{0.15cm}\\
\textit{Remark:} Given a continuously differentiable family, $\big\{H_t\,\big|\,0\leq t\leq 1\big\}$, of matrices on $\mathbb{C}^{N}$
with the properties that $H_0$ is a selfadjoint, diagonal operator, one can find the eigenvalues $E_t^0 \leq E_t^1\leq \dots \leq E_t^{N-1}$ 
and the corresponding eigenprojections $\Pi_t^0, \Pi_t^1, \dots, \Pi_t^{N-1}$, for $0<t\leq 1$, by solving an initial-value problem 
consisting of a system of coupled non-linear ordinary differential equations for these quantities. These equations can be derived 
by using \eqref{PT}. Some details concerning this so-called \textit{``infinitesimal perturbation theory''} are discussed in Appendix A.

Next, we apply infinitesimal perturbation theory, in particular \eqref{PT}, to equations \eqref{mixing states} and \eqref{convex combi}
in order to derive expressions for the quantities $p^{\delta}[t, t+dt]$ and $\Pi_{t+dt}^{\delta},$ \mbox{$ \delta =0,1, \dots, N-1$,}
(with $p^0 \equiv p_{nj}$). For this purpose, we set $H_0:= \Pi_t, V:= \mathfrak{L}[\Pi_t]$, and $\varepsilon:= dt$. For these choices, 
the spectrum of $H_0$ is given by $\big\{1, 0, \dots,0 \big\}$, the eigenvalue 1 being simple and the eigenvalue 0 being $(N-1)$-fold 
degenerate; (recall that $\Pi_t$ is a rank-1 orthogonal projection).
In the present context, the operator $\mathcal{S}$ introduced in \eqref{generator} is given by
\begin{equation}\label{S}
\mathcal{S}\equiv \mathcal{S}_t:= -\Pi_t^{\perp}\,\mathfrak{L}_{\alpha}[\Pi_t]\, \Pi_t + \Pi_t\,\mathfrak{L}_{\alpha}[\Pi_t]\, \Pi_t^{\perp}\,.
\end{equation}
Equation \eqref{convex combi} and the first equation in \eqref{PT} (sometimes called Feynman-Hellmann theorem) 
then imply that the probability, $p_{nj}$, that there is no ``quantum jump'' during the time interval $[t, t+dt)$ is given by 
\begin{equation}\label{p-nj}
p_{nj}[t, t+dt] = 1+\text{Tr}\big(\Pi_t\, \mathfrak{L}_{\alpha}[\Pi_t]\big)\, dt + \mathcal{O}(dt^2)\,,
\end{equation}
and, for the state, $\Pi_{t+dt}$, of the atom at time $t+dt$ corresponding to the eigenvalue $p_{nj}[t, t+dt]$, 
the second equation in \eqref{PT} yields
\begin{equation}\label{state of atom}
\Pi_{t+dt} = \Pi_t + \Big\{\Pi_t^{\perp}\,\mathfrak{L}_{\alpha}[\Pi_t]\,\Pi_t + \Pi_t\,\mathfrak{L}_{\alpha}[\Pi_t]\,\Pi_t ^{\perp}\Big\} dt + \mathcal{O}(dt^2)\,.
\end{equation}
Using the remark following Eq.~\eqref{PT}, we conclude that the probabilities $p^{\delta}[t,t+dt], \delta =1, \dots, N-1,$ 
for observing a ``quantum jump'' during the time interval $[t, t+dt)$ are given by the eigenvalues of the \textit{non-negative} matrix 
$\Pi_t^{\perp} \,\mathfrak{L}_{\alpha}[\Pi_t]\, \Pi_t^{\perp}\cdot dt \overset{\eqref{Lindblad-1}}{=} 
\alpha \sum_{(ij)\in \mathcal{T}} \Pi_t^{\perp}\,D_{ij}\, \Pi_t  \,D_{ij}^{*}\,\Pi_t^{\perp}\, dt$; i.e.,
\begin{equation}\label{p-jump}
\text{spec}\big(\Pi_t^{\perp} \,\mathfrak{L}_{\alpha}[\Pi_t]\, \Pi_t^{\perp}\, dt \big)= \big\{p^{\delta}[t, t+dt]\,\big|\, \delta=1, \dots, N-1\big\}\,,
\end{equation}
up to corrections of order $\mathcal{O}(dt^2)$, where $p^{\delta}[t, t+dt] \geq 0, \,\, \forall \,\, \delta=1, \dots, N-1$.

From \eqref{p-nj} and \eqref{state of atom} we derive differential equations for the probability $p_{nj}[t_1, t_2]$
that no ``quantum jump'' occurs during the time interval $[t_1, t_2)$ and for the state, $\Pi_t$, of the atom for $t\in [t_1, t_2)$, 
(i.e., in the absence of ``quantum jumps'').
\begin{enumerate}
\item[(I)]{From \eqref{p-nj} we get
\begin{align}\label{Feynman-Hellmann}
\begin{split}
\frac{d \ln (p_{nj}[t, t+dt])}{dt} &= \text{Tr}\big(\Pi_t\, \mathfrak{L}_{\alpha}[\Pi_t]\big) = -\alpha \sum_{(ij)\in \mathcal{T}} 
\text{Tr}\big(\Pi_t^{\perp}\,D_{ij}\, \Pi_t  \,D_{ij}^{*}\,\Pi_t^{\perp}\big) <0\, \quad \Rightarrow\\
p_{nj}[t_1, t_2]&= \text{exp}\Big\{\int_{t_1}^{t_2} \text{Tr}\big(\Pi_t\, \mathfrak{L}_{\alpha}[\Pi_t]\big) dt\Big\} < 1\,,
\end{split}
\end{align}
where we have used that 
\begin{align*}
\text{Tr}\big(\Pi_t \,\mathfrak{L}_{\alpha}[\Pi_t]\big)&= \underbrace{\text{Tr}\big(\mathfrak{L}_{\alpha}[\Pi_t]\big)}_{=0} - 
\text{Tr}\big(\Pi_t^{\perp}\, \mathfrak{L}_{\alpha}[\Pi_t]\big)\\
&= -\alpha \sum_{(ij)\in \mathcal{T}} \text{Tr}\big(\Pi_t^{\perp}\,D_{ij}\, \Pi_t  \,D_{ij}^{*}\,\Pi_t^{\perp}\big)\,<0\,.
\end{align*}
}
\item[(II)]{From \eqref{state of atom} we derive the following differential equation for the state $\Pi_t$ of the atom in the absence 
of ``quantum jumps'' (meaning that the eigenprojection corresponding to the eigenvalue \mbox{$p_{nj}[t,t+dt]$} is chosen for all times 
$t$ in a time interval $[t_1, t_2)$):
\begin{equation}\label{time-dep}
\frac{d\Pi_t}{dt}= \Pi_t^{\perp}\, \mathfrak{L}_{\alpha}[\Pi_t]\, \Pi_t + \Pi_t\, \mathfrak{L}_{\alpha}[\Pi_t]\, \Pi_t^{\perp}\,, 
\qquad  t_1\leq t< t_2\,.
\end{equation}
This is a system of \textit{non-linear (cubic) differential equations} for the state of the atom during a time interval $[t_1, t_2)$
without ``quantum jumps.''
}
\item[(III)]{Combining \eqref{p-nj} with \eqref{p-jump} and using the cyclicity of the trace, we find that
\begin{equation}\label{sum rule}
p_{nj}[t, t+dt] + \sum_{\delta=1}^{N-1} p^{\delta}[t,t+dt] = 1 + \text{Tr}\big( \mathfrak{L}_{\alpha}[\Pi_t]\big) = 1\,,
\end{equation}
as expected.
}
\end{enumerate}
It may be instructive to specialize formulae \eqref{Feynman-Hellmann} and \eqref{time-dep} to the case where $\alpha=0$, i.e.,
where the atom is decoupled from the radiation field. Then
$$\frac{\ln (p_{nj}[t, t+dt])}{dt} = \text{Tr}\big(\Pi_t \,\mathfrak{L}_{0}[\Pi_t]\big)= -i\text{ Tr}\big(\Pi_t \,[H_A, \Pi_t]\big) \equiv 0\,,$$
hence 
$$p_{nj}[t_1,t_2]\equiv 1, \,\,t_1<t_2 \text{ arbitrary},\quad\text{and}\quad p^{\delta}[t, t+dt] = 0,\,\,\forall\,\, \delta=1,\dots, N-1.$$
Furthemore,
\begin{align*}
\frac{d\Pi_t}{dt}=&  \Pi_t^{\perp}\, \mathfrak{L}_{0}[\Pi_t]\, \Pi_t + \Pi_t\, \mathfrak{L}_{0}[\Pi_t]\, \Pi_t^{\perp}\\
=& -i\big\{\Pi_t^{\perp}\,[H_A, \Pi_t] \,\Pi_t + \Pi_t \,[H_A, \Pi_t]\, \Pi_t^{\perp}\big\} = -i[H_A, \Pi_t]\,,
\end{align*}
i.e., for $\alpha = 0$, the time-evolution of the state of the atom is described by the Schr\"odinger-von Neumann equation 
with Hamiltonian $H_A$, as one should expect.

\subsection{A measure on the space of atomic state trajectories with ``quantum jumps''}
Next, we construct an analogue of the measure, $W_{\omega}$, on trajectories of random walkers on $\mathbb{Z}^{\nu}$
given in Eq.~\eqref{proba} of Sect.~3. We suppose that the state of the atom exhibits ``quantum jumps'' at random times
$t_1<\dots < t_n$, with $t_1>\underline{t}=0, t_n< \overline{t}$ and that, during time intervals $[t_j, t_{j+1})$ between the
$j^{th}$ and the $j+1^{st}$ quantum jump, the state of the atom evolves continuously according to Eq.~\eqref{time-dep}. This state
is denoted by $\Pi_t^{\delta_{j-1} |\delta_{j}}$, with $j=1, \dots, n,\, (\delta_0:=1)$; the initial condition at $t=\underline{t}=0$ is given by
a rank-1 orthogonal projection $\Pi_0^{0|1}:= \Pi^{0}$. This state gives rise to a trajectory $\Pi_t^{0|1}$ solving 
Eq.~\eqref{time-dep} for $t\in[0, t_1)$. In the interval $[t_1, t_1 + dt_1)$ the state of the atom performs a ``quantum jump'' 
to a state denoted by $\Pi_{t_1}^{\delta_0|\delta_1}$, which is the eigenstate of the matrix 
$$\big[\Pi_{t_1}^{0|\delta_0}\big]^{\perp} \,\mathfrak{L}_{\alpha}[\Pi_{t_1}^{0|\delta_0}]\, [\Pi_{t_1}^{0|\delta_0}\big]^{\perp} dt_1$$
corresponding to the eigenvalue $p^{\delta_1}[t_1, t_1 + dt_1]$; here we adopt the convention for the choice of the label $\delta_1$ 
specified in \eqref{probabilities} (and, for simplicity, we assume that there are no degeneracies among the eigenvalues $p^{\delta}$). 
Subsequently, one solves Eq.~\eqref{time-dep} for a state $\Pi_{t}^{\delta_0|\delta_1}$, for $t_1\leq t<t_2$, using $\Pi_{t_1}^{\delta_0|\delta_1}$ 
as an initial condition at time $t=t_1$. The probability for the state of the atom \textit{not} to perform a ``quantum jump'' 
during the time interval $[t_1, t_2)$ is given by $p_{nj}^{\delta_0|\delta_1}[t_1, t_2]$ given in Eq.~\eqref{Feynman-Hellmann}, 
with $\Pi_t \mapsto \Pi_t^{\delta_0|\delta_1}$.
Let $1\leq j \leq n$; the state of the atom at times $t\in [t_j, t_{j+1})$ is denoted by $\Pi_t^{\delta_{j-1}|\delta_j}$. It solves the 
differential equation \eqref{time-dep} with initial condition at $t=t_j$ given by $\Pi_{t_j}^{\delta_{j-1}|\delta_j}$. The probability 
not to perform a ``quantum jump'' in the time interval $[t_j, t_{j+1})$ is given by
\begin{equation}\label{no jump}
p_{nj}^{\delta_{j-1}|\delta_j}[t_j, t_{j+1}]=\text{exp}\Big\{\int_{t_j}^{t_{j+1}} \text{Tr}\big(\Pi_t^{\delta_{j-1}|\delta_j}\, 
\mathfrak{L}_{\alpha}[\pi_t^{\delta_{j-1}|\delta_j}]\big)dt\Big\}\,.
\end{equation}
In the interval $[t_{j+1}, t_{j+1}+dt_{j+1})$ the atom performs a ``quantum jump'' from the state $\Pi_{t_{j+1}}^{\delta_{j-1}|\delta_j}$ 
to a new state $\Pi_{t_{j+1}}^{\delta_{j}|\delta_{j+1}}$ with probability given by $p^{\delta_{j+1}}[t_{j+1}, t_{j+1}+dt_{j+1}]$, where
\begin{align}\label{jump proba}
\begin{split}
&\Pi_{t_{j+1}}^{\delta_{j}|\delta_{j+1}} \text {is the eigenprojection of the matrix }\,\,
\big[\Pi_{t_{j+1}}^{\delta_{j-1}|\delta_j}\big]^{\perp} \,\mathfrak{L}_{\alpha}[\Pi_{t_{j+1}}^{\delta_{j-1}|\delta_j}]\, 
\Pi_{t_{j+1}}^{\delta_{j-1}|\delta_j}\big]^{\perp} dt_{j+1}\\
&\text{corresponding to the eigenvalue }\,\,p^{\delta_{j+1}}[t_{j+1}, t_{j+1}+dt_{j+1}]\,,
\end{split}
\end{align}
the convention for the choice of the label $\delta_j$ being given by \eqref{probabilities}. Note that, just for simplicity, 
we assume that all the eigenvalues $p^{1}[t_{j+1}, t_{j+1}+dt_{j+1}],\dots, p^{N-1}[t_{j+1}, t_{j+1}+dt_{j+1}]$, arranged
in decreasing order as in \eqref{probabilities}, are \textit{simple}. This assumption can be eliminated.

A \textit{trajectory of states} of an atom exhibiting $n$ ``quantum jumps'' is denoted by $\mathfrak{T}_n$; and we set
 $\mathfrak{T}:= \bigcup_{n=0}^{\infty} \mathfrak{T}_n$. If the evolution is monitored during a specified interval of times $[0, \overline{t}]$ we
 write $\mathfrak{T}_n[0, \overline{t}],$ etc. We propose to calculate the \textit{probability} of a trajectory (or \textit{``history''}),
 $\mathfrak{T}_n[0, \overline{t}]$, of states given by
\begin{equation}\label{trajectory}
\mathfrak{T}_n \equiv \mathfrak{T}_n[0, \overline{t}\,] := \Big\{ \Pi_t^{\delta_{j-1}|\delta_j}\Big| t_j< t < t_{j+1},\, j=0,1, \dots, n\,\Big\}\,,
\end{equation}
with $t_{n+1}:= \overline{t}$, assuming that the $j^{th}$ quantum jump occurs in the time interval $[t_j, t_j + dt_j)$:
\begin{align}\label{Poisson measure}
\text{prob}(\mathfrak{T}_n[0, \overline{t}])\equiv \,&W_{\mathfrak{T}_n}\big[\delta_1, t_1, \dots, \delta_n, t_n; 0, \overline{t} \big] \prod_{j=1}^{n} dt_j =\nonumber\\
&=\Big\{ \prod_{j=0}^{n-1} p_{nj}^{\delta_{j-1}| \delta_j}[t_j, t_{j+1}]\, p^{\delta_{j+1}}[t_{j+1}, t_{j+1}+ dt_{j+1}]\Big\} p_{nj}^{\delta_{n-1}|\delta_n}[t_n, \overline{t}\,]\,,
\end{align}
with $\delta_{-1} :=0, \, \delta_0:=1.$ This formula enables one to determine the probabilities of measureable sets in the path space, 
$\Xi[0, \overline{t}\,]$, of histories, $\mathfrak{T}[0, \overline{t}\,]$, of \textit{individual} atoms performing an arbitrary 
number of ``quantum jumps'' in the time interval $[0, \overline{t}\,]$, for an arbitrary $\overline{t}>0$. 
Neglecting (eigenvalue) degeneracies, the path space $\Xi$ can be defined as
$$\Xi[0, \overline{t}\,]:= {\bigtimes}_{t\in [0, \overline{t}\,]} \dot{\mathfrak{P}}_t \,,$$
where $\mathfrak{P}$ is complex projective space $\mathbb{C}P^{N-1}$, and $\dot{\mathfrak{P}}$ is the one-point compactification of
$\mathfrak{P}$. The stochastic time evolution of states of atoms described here is called a \textbf{``quantum Poisson process''}.

\textit{Remark:} If one takes an average, $\mathbb{E}$, over all histories of state trajectories $\mathfrak{T}[0, \overline{t}\,]$,
weighted according to the measure introduced in \eqref{Poisson measure}, then one recovers the Lindblad evolution of ensemble 
states given in \eqref{Lindblad} and \eqref{Lindblad-1}; i.e.,
\begin{equation}\label{history sum}
\mathbb{E}\big[\mathfrak{T} [0, \overline{t}\,]\big] =\big\{\Omega_t\,\big|\, 0\leq t\leq \overline{t}, \Omega_0=\Pi_0^{0}\equiv \Pi_0^{0|1}\big\},
\end{equation}
where $\Omega_t$ is a solution of the Lindblad equation \eqref{Lindblad}. Equation \eqref{history sum} 
follows from the ``sum rule'' in \eqref{sum rule}. In words, an ensemble state at time $\overline{t}$ is given by a weighted
sum over \textit{``histories''} of states of individual systems, all prepared in the same initial state.

\section{Stochastic Quantum Evolution \textit{With} Photon Detection}\label{Q-Poisson}
In this section we describe the stochastic dynamics of individual atoms spontaneously emitting photons, assuming that these
photons will be detected by a photomultiplier, as described in Sect.~\ref{Physical systems}. The interaction Hamiltonian $H_I$ introduced
in \eqref{interaction Ham} describes processes where an atom in a state $\Psi$ overlapping with the eigenstate $\psi_j$ 
of the atomic Hamiltonian $H_A$ jumps to an eigenstate $\psi_i$ of $H_A$ and emits a photon in a state $\xi_{ij}$. The initial state 
of the system at a time $t$ right \textit{before} the emission process is given by
\begin{equation}\label{in-state}
\Psi_t\otimes \big|\emptyset, t\big> \otimes \big|s\big>\,,
\end{equation}
where $\Psi_t$ is the state vector of the atom right before spontaneous emission at time $t$, $\big|\emptyset, t\big>$ 
is the vacuum in the Fock space $\mathcal{F}_{\geq t}$, and $\big|s\big>$ is the silent state of the photomultiplier; 
see Sect.~\ref{Physical systems}. Right \textit{after} an emission process occurring in the time interval $[t, t+dt)$, 
the state vector of the system is given by
\begin{equation}\label{out-state}
\psi_i\otimes  \int_t^{t+dt}a^{*}(\tau, \xi_{ij})\,d\tau\,\big|\emptyset, t\big> \otimes \big|\{\xi_{ij}\}\big>\,,
\end{equation}
where a (one-photon) state $\xi_{ij}$ has been filled with a photon, and the photomultiplier at time $t+dt$ is in an excited state 
$\big|\{\xi_{ij}\}\big>$. In Sect.~\ref{Physical systems} we have assumed that the (improper) states of the photomultiplier
given by
\begin{equation}\label{orthogonality}
 \big|s\big>,\,\,\, \big|\{\xi_{ij}\}\big>\,\, \text{ and }\,\,\, \big|\{\xi_{k\ell}\}\big>, 
\end{equation} 
are \textit{orthogonal} to one another whenever $i \not= k$, with $(ij)$ and $(k\ell)$ 
belonging to the set $\mathcal{T}$ of allowed atomic transitions. Furthermore,
\begin{equation}\label{orthogonality-1}
a^{*}(\cdot, \xi_{ij})\big|\emptyset \big>\,\, \text{ and }\,\, a^{*}(\cdot, \xi_{k\ell})\big|\emptyset \big>
\end{equation}
are orthogonal to one another whenever $i \not= k$, with $(ij), (k\ell) \text{ in } \mathcal{T}$, $ j>i$ and $ \ell>k$;
see the \textit{simplifying assumptions} after Eq.~\eqref{interaction Ham} in Sect.~2.1.

When applying the state-reduction postulate of the $ETH$ - Approach (see \cite{BFS, FGP}) in the situation discussed
here one must keep track of the fact that the states of the photons emitted by the atom satisfy the orthogonality properties 
described in \eqref{orthogonality-1} and that the state of the atom after spontaneous emission of a photon is entangled with
the state of the emitted photon  -- in contrast to the situation discussed in Sect.~4, where the photons emitted by the 
atom do not appear explicitly when applying the state-reduction postulate.
Yet, since, in our models, the velocity of light is assumed to be $\infty$, photons emitted by the atom and the 
photomultiplier \textit{immediately} disappear from the system, so that, right after a spontaneous emission of a 
photon by the atom and the firing of the photomultiplier, the radiation field is again found
in its vacuum state and the state of the photomultiplier is the silent state $\big|s\big>\big< s \big|$. 
As in Sect.~4, this implies that the stochastic evolution of the state of the system can be encoded completely in an
\textit{effective time evolution} of the state of the atom, which, however, differs from the one described in Sect.~4. 
Let $\Pi_t=\big|\Psi_t\big>\big< \Psi_t\big|$ be the state of the atom at time $t$. We conclude that, in the present 
situation, the alternatives between different states of the atom at time $t+dt$ imposed by a correct application of the 
state-reduction postulate of the $ETH$ - Approach are given by
\begin{equation}\label{atomic state}
\Pi_{t+dt} = \begin{cases}
p_{nj}[t, t+dt; \Pi_t]^{-1} \Big\{\Pi_t + \mathfrak{L}_{\alpha}'[\Pi_t] dt \Big\}, &\text{ with probability }\,\,p_{nj}[t, t+dt; \Pi_t]\,,\\
\big|\psi_i\big>\big<\psi_i\big|, & \text{ with probability }\,\, p^{i}[t, t+dt; \Pi_t]\,,\end{cases}
\end{equation}
 i<N-1\,, where $\mathfrak{L}_{\alpha}'[\cdot]$ has been defined in \eqref{21'}, $\psi_i$ is an eigenstate of the atomic 
 Hamiltonian $H_A$ (see \eqref{atomic Ham}), and
\begin{align}\label{jumping prob}
\begin{split}
p_{nj}[t, t+dt, \Pi]:=& 1+ \text{Tr}\big( \Pi\, \mathfrak{L}_{\alpha}'[\Pi]\big)\,dt = 1-\alpha \sum_{(ij)\in \mathcal{T}, j>i} \text{Tr}\big(D_{ij}\Pi D_{ij}^{*}\big)\, 
dt< 1\,,\\
p^{i}[t, t+dt; \Pi]:=& \sum_{j\,:\, (ij)\in \mathcal{T},\, j>i} \alpha \text{Tr}\big(D_{ij}\Pi D_{ij}^{*}\big)\,dt\,.
\end{split}
\end{align}

It may be appropriate to sketch the derivation of these formulae. We first consider the evolution equation 
\begin{equation}\label{(21')}
\dot{\Omega}_t = \mathfrak{L}'[\Omega_t]\,, \quad \text{with }\,\, \,\mathfrak{L}' [\Omega]:= -i[H_A, \Omega] -
\frac{1}{2}\big\{\Omega, D\big\}\,, \,\,\, D>0\,,
\end{equation}
where, in the models introduced in Sect.~\ref{Physical systems}, $\mathfrak{L}'= \mathfrak{L}'_{\alpha}$ and 
$D:= \alpha \sum_{(ij)\in \mathcal{T}:j>i} D_{ij}^{*} D_{ij}$.
Suppose that $\Omega_t$ solves Eq.~\eqref{(21')} with initial condition $\Omega_{t=\underline{t}}= \Pi_{\underline{t}}$,
where $\Pi_{\underline{t}}$ is a rank-1 orthogonal projection. Then $\Omega_{t}$ is proportional (but \textbf{not} equal) to a rank-1 
orthogonal projection,  denoted by $\Pi_{t}$, at all times $t > \underline{t}$, with
\begin{equation}\label{rescaling}
\Pi_t= p_{nj}[0, t]^{-1} \Omega_t\,, \quad \text{where }\,\, p_{nj}[0,t] = \text{Tr}\Big(\text{exp}\big(t \mathfrak{L}'\big)[\Pi_{\underline{t}=0}]\Big)<1\,.
\end{equation}
This is a general property of solutions 
of evolution equations on the space of trace-class operators of the form of \eqref{(21')}, with $D\geq 0$, as is 
verified by noting that the evolution given by Eq.~\eqref{(21')} does not preserve the trace; 
in fact, the trace of a solution $\Omega_t$ is decreasing in $t$, as follows from the inequality
\begin{align*}
\text{tr}\big(\Omega_{t+dt}\big) = &\text{Tr}\big(\Omega_t\big)  - \text{Tr}\big(i[H, \Omega_t] + \Omega_t D\big)\, dt\\
=& \text{Tr}\big(\Omega_t\big) -  \underbrace{\text{Tr}\big(\Omega_t \,D\big)\, dt}_{> 0}< \text{Tr}\big(\Omega_t\big)\,,
 \end{align*}
where we have used that $D\geq 0$; and corrections of $\mathcal{O}(dt^{2})$ can be ignored, as $dt \rightarrow 0$.

During every time interval $[t, t+dt)$, the state, $\Pi_t$, of an atom \textit{either} does \textit{not} make a jump and evolves according to
the first equation in \eqref{atomic state} (see also \eqref{(21')} and \eqref{rescaling}); \textit{or} it spontaneously emits a photon and 
makes a jump to a state in the range of one of the operators $D_{ij}, (ij)\in \mathcal{T}$, corresponding to an eigenvector, $\psi_i$, 
of the atomic Hamiltonian $H_A$. Because of the orthogonality properties in \eqref{orthogonality} and \eqref{orthogonality-1} 
these are mutually exclusive events, and applying the state-reduction postulate of the \mbox{$ETH$ -} Approach, 
we conclude that the equations in \eqref{atomic state} must hold. The probability of \textit{not} making a jump is close to 1, namely 
\begin{align}\label{no-jump prob}
\begin{split}
p_{nj}[t, t+dt; \Pi_t]  \overset{\eqref{(21')}}{=}& \text{Tr}\big(\Pi_t + \mathfrak{L}_{\alpha}'[\Pi_t ]dt\big)
= 1 - \text{Tr}\big(\Pi_t\, D\big)\,dt\\
\overset{\eqref{21'}}{=}& 1- \alpha \sum_{(ij)\in \mathcal{T}} \text{Tr}\big(D_{ij}\Pi_t D_{ij}^{*}\big)\,dt\,.
\end{split}
\end{align}
The probability for spontaneously emitting a photon during a time interval $[t, t+dt)$ is proportional to $dt$; it is given by
the right side of the second equation in \eqref{jumping prob}. This equation implies that
$$p_{nj}[t, t+dt; \Pi_t] + \sum_{i}p^{i}[t, t+dt; \Pi_t] =1,$$
as expected. 

The unraveling of the Lindblad equation described in this section is analogous to the one described in \cite{BCFFS}.

\subsection{A measure on state trajectories with ``quantum jumps'' when emitted photons are detected}
Next, we construct a formula for a measure on the space of trajectories of atomic states analogous to the one presented in \eqref{Poisson measure}.
We consider a trajectory, $\mathfrak{T}_{n}[0, \overline{t}\,]$, of states of an atom spontaneously emitting photons at $n$ randomly
chosen times $\underline{t}\equiv 0<t_1<\dots< t_n< \overline{t}$, for $n=0,1,2,\dots$
It is easy to integrate Eq.~\eqref{no-jump prob} for $p_{nj}$; as claimed in \eqref{rescaling}, one finds that
\begin{equation}\label{nj prob}
p_{nj}[t_1, t_2; \Pi] = \text{Tr}\Big(e^{(t_2 - t_1)\, \mathfrak{L}'}\big[\Pi\big]\Big)\,.
\end{equation}
A formula for the probabilities $p^i[t, t+dt; \Pi]$ has been given in \eqref{jumping prob}. 
We consider a trajectory, $\mathfrak{T}_n$, of states of the atom 
\begin{equation}\label{atom traj}
\mathfrak{T}_n[0, \overline{t}\,] :=\Big\{\Pi^{i_j}_t\,\Big|\, \Pi^{i_j}_t :=\frac{\text{exp}\big\{(t-t_j)\mathfrak{L}_{\alpha}'\big\}\big[\Pi^{i_j}\big]}
{\text{Tr}\Big(\text{exp}\big\{(t-t_j)\mathfrak{L}_{\alpha}'\big\}\big[\Pi^{i_j}\big]\Big)}, \,t_j < t < t_{j+1},\, j=0,1, \dots, n\Big\}\,,
\end{equation}
where $\Pi^{i}:= \big| \psi_{i}\big>  \big< \psi_{i}\big| $, and $\psi_i$ is an eigenstate of $H_A$, except that, at time $t= t_0 \equiv \underline{t}=0$, $\Pi^{i_0}:= \Pi$, where $\Pi$
is the initial state of the atom, which is an arbitrary pure state, and $t_{n+1}=\overline{t}$ is the time when the recording of the 
trajectory ends.

The probability density of the trajectory $\mathfrak{T}_n[0, \overline{t}\,]$ with respect to the Lebesgque measure $dt_1 \dots dt_n$ 
is given by
\begin{align}\label{Poisson-1}
\begin{split}
W_{\mathfrak{T}_{n}[0, \overline{t}\,]}  \big[i_1, &t_1, \dots, i_n, t_n ; \Pi \big] \prod_{j=1}^{n} dt_j\\
&= \Big\{\prod_{j=0}^{n-1} p_{nj} [t_j, t_{j +1}; \Pi^{i_j}] \, p^{i_{j+1}} \big[t_{j+1}, t_{j+1} + dt_{j+1}; \Pi^{i_j}_{t_{j+1}} \big]
\Big\} p_{nj}[t_n, \overline{t}; \Pi^{i_n}]\,,
\end{split}
\end{align}
where the probabilities $p^{i}[t, t+dt; \cdot\,]$ are defined in \eqref{jumping prob}, for $n=0,1,2, \dots$

As in Sect.~4, Eq.~\eqref{history sum}, one verifies that the \textit{sum over all histories}  of atomic states between an initial time 
$\underline{t}=0$ and a final time $\overline{t}$, weighted according to the measure introduced in \eqref{Poisson-1}, reproduces 
the Lindblad evolution of \textit{ensemble states}. This is proven with the help of the Dyson series expansion applied to 
$\text{exp}\big\{\overline{t}\, \mathfrak{L}_{\alpha}\big\}\big[\Pi\big]$, treating the terms 
$$\alpha\,\sum_{j: (ij)\in \mathcal{T},\, j>i}D_{ij} \Pi D_{ij}^{*}, \quad i=0, \dots, N-2,$$
as perturbations and $\mathfrak{L}_{\alpha}'$ as the unperturbed generator of the evolution; (see \cite{BCFFS} for
similar results).

\textit{Remark:} Using the results concerning \textit{``relaxation to the groundstate''} mentioned in Subsect.~2.3, one sees
that Equation \eqref{history sum} and the result just described imply that, under the assumption that the radiation field is 
prepared in the vacuum state, the state of an arbitrary atom ends up approaching the groundstate $\psi_0$, as time 
$\overline{t}$ tends to $\infty$ (independently of whether photons emitted by the atom are recorded or not). Of course, 
this conclusion does \textit{not} hold if the radiation field is prepared in a state containing photons at all times or if the 
atom exhibits never-ending Rabi oscillations.

\section{Two-level atoms}\label{Example}
In this section we consider physical systems consisting of two-level atoms, i.e., $N=2$, interacting with the quantized radiation
field in the limiting regime where the velocity of light is infinite. Our interest in this special case is motivated by the desire to
present some strikingly explicit results, which, we expect, can be used to make predictions that could be verified in experiments.
In particular, the results in subsection 6.1 might lead to experimentally verifiable predictions, and the dependence of atomic
state trajectories on whether photons emitted by the atom are recorded, or not, might also point to experimental signatures.

The Hilbert space of a two-level atom is given by 
$\mathfrak{h}_A= \mathbb{C}^{2}$, and its Hamiltonian is chosen to be
\begin{equation}\label{H-A}
H_A:= \hbar \omega \begin{pmatrix} 1&0\\0&0 \end{pmatrix}\,, \qquad \omega>0\,.
\end{equation}
To streamline our notations we henceforth set $\hbar = \omega =1$. 
States of a two-level atom are given by non-negative $2\times2$ matrices $\Omega$, with $\text{Tr}\big(\Omega\big) = 1$.
We require all the assumptions stated in Subsects.~2.2 and 2.3; in particular, the initial state of the radiation field is the vacuum
state. It then follows that the dynamics of such a system can be completely encoded in the \textit{effective dynamics} of the state of the
atom. The dynamics of an average of the states over a large ensemble of identical systems (which we have called ``ensemble states''), 
all prepared in the same initial state $\Omega_0 \otimes \big|\emptyset\big>\big<\emptyset\big|$, is described by an effective 
evolution equation for the state $\Omega_t, t\geq \underline{t}:=0,$ of the atom; the state of the radiation field being the vacuum state, 
$\big|\emptyset,t\big>\big< \emptyset, t\big|$, at \textit{all} later times $t$. For the systems discussed in this section, 
the time evolution of ensemble states is described by the Lindblad equation
\begin{align}\label{emission}
\begin{split}
&\dot{\Omega}_t = \mathfrak{L}_{\alpha}\big[\Omega_t\big]\,, \qquad \text{where }\\
\mathfrak{L}_{\alpha}\big[\Omega\big]:= &-i\big[H_A, \Omega\big] + \alpha\big[\sigma_{-}\,\Omega\, \sigma_{+} 
-\frac{1}{2} \big\{\Omega, \sigma_{+}\sigma_{-}\big\}\big]\,.
\end{split}
\end{align}
The ``raising- and lowering operators,'' $\sigma_{\pm},$ are given by
$$\sigma_{+}:= \begin{pmatrix} 0&1\\0&0\end{pmatrix}\,, \qquad \sigma_{-}:= \begin{pmatrix} 0&0\\1&0 \end{pmatrix}\,,$$
respectively.

For two-level atoms, it is convenient to parametrize density matrices by vectors in $\mathbb{R}^{3}$ of length $\leq 1$, namely
\begin{equation}\label{Bloch}
\Omega\equiv \Omega(\vec{n}):= \frac{1}{2}\Big(\mathbf{1} + \vec{n}\cdot \vec{\sigma}\Big), \quad \vec{n}\in \mathbb{R}^{3}, \,\,\,\text{ with }\,\,\,
\big|\vec{n}\big| \leq 1\,,
\end{equation}
where $\vec{\sigma}:=\big(\sigma_1, \sigma_2, \sigma_3\big)$ is the vector of Pauli matrices (i.e., $\sigma_1= \sigma_{+}+ \sigma_{-}$,
$\sigma_2=-i\sigma_{+} + i \sigma_{-}$, and $\sigma_3 = 2\sigma_{+}\,\sigma_{-} - \mathbf{1}$). The state 
$\Omega(\vec{n})$ is \textit{pure} iff $\vec{n}$ is a unit vector, i.e., if $\vec{n}$ lies on the so-called \textit{Bloch sphere}; moreover,
$$\Omega(\vec{n})+ \Omega(- \vec{n}) = \mathbf{1}\,, \quad \text{and if }\, |\vec{n}|=1 \,\text{ then }\, \Omega(\vec{n})\cdot \Omega(-\vec{n})=0\,.$$
One may now re-write the evolution equation \eqref{emission} as an equation for the vector $\vec{n}_t$ parametrizing the state
$\Omega_t\equiv \Omega(\vec{n}_t)$ at time $t$, which is given by
\begin{equation}\label{Eq for n}
\dot{\vec{n}}(t)= \underbrace{ \vec{e}_3 \wedge \vec{n}(t)}_{\text{precession around } \vec{e}_3} 
\underbrace{-\frac{\alpha}{4}\Big[2\vec{e}_3 + \vec{n}(t) + n_3(t)\cdot\vec{e}_3\Big]}_{\text{dissipative terms}},
\end{equation}
where $\big\{\vec{e}_1, \vec{e}_2, \vec{e}_3\big\}$ is the standard orthonormal basis in $\mathbb{R}^{3}$.
This equation implies that
\begin{equation}\label{relax}
\vec{n}(t)\rightarrow -\vec{e}_3, \,\,\text{ as }\,t\rightarrow \infty, \qquad \text{with }\,\,\, 
\Omega(-\vec{e}_3)= \begin{pmatrix} 0&0 \\0& 1\end{pmatrix}\,,
\end{equation}
i.e., $\Omega(-\vec{e}_3)$ is the groundstate of $H_A$; (compare to \eqref{R to gs}).

Next, we ``unravel'' Eq.~\eqref{emission} or, equivalently, \eqref{Eq for n} by specializing the methods developed in Sects.~\ref{Unraveling Lindblad}
and \ref{Q-Poisson} to the present example. We first assume that photons emitted by the atom are not detected, following the procedure in
Sect.~\ref{Unraveling Lindblad}. From now on, we write unit vectors on the Bloch sphere as $\vec{e}=\big(e_1, e_2, e_3\big)$, with 
$e_1^2+e_2^2 +e_3^2 =1$, and vectors in the interior of the Bloch sphere as $\vec{n}$. (Note that $e_i$ is the i-component 
of a unit vector $\vec{e}\in \mathbb{R}^{3}$, while $\vec{e}_{i}$ is the $i^{th}$ vector in the standard orthonormal basis of 
$\mathbb{R}^{3}$.) We assume that, at some time $t$, the atom is prepared in a \textit{pure} state, i.e., 
$\Omega_t= \Omega\big(\vec{e}(t)\big), $with $\big|\vec{e}(t)\big|=1$.  Then the ensemble state of an atom at time $t+dt$ 
($dt$ small) is given by $\Omega\big(\vec{n}(t+dt)\big)$, where
\begin{equation}\label{n(t+dt)}
\vec{n}(t+dt)= \vec{e}(t)+\Big\{\vec{e}_3 \wedge \vec{e}(t) - \frac{\alpha}{4}\big[2\vec{e}_3 + \vec{e}(t) + e_{3}(t)\cdot \vec{e}_3\big]\Big\} dt 
+ \mathcal{O}(dt^{2})\,.
\end{equation}
The principles of the $ETH$ - Approach to Quantum Mechanics then imply that the state of an \textit{individual} atom at time $t+dt$
corresponds to a \textit{unit vector} $\vec{e}(t+dt)$ given by
\begin{align}\label{purification}
\vec{e}(t+dt)=\begin{cases}\,\,\,\, \frac{\vec{n}(t+dt)}{\big|\vec{n}(t+dt)\big|}, &\quad \text{with probability }\,\,p_{nj}[t, t+dt]\,,\\
- \frac{\vec{n}(t+dt)}{\big|\vec{n}(t+dt)\big|}, &\quad \text{with probability }\,\,p_{flip}[t, t+dt]\,,
\end{cases}
\end{align}
where
\begin{align}\label{flips}
\begin{split}
p_{nj}[t, t+dt] &= \frac{1+ \big|\vec{n}(t+dt)\big|}{2} = 1- \mathcal{O}(dt)\,,\\
p_{flip}[t, t+dt] &= \frac{1- \big|\vec{n}(t+dt)\big|}{2} = \mathcal{O}(dt)\,,
\end{split}
\end{align}
so that $p_{nj}[t,t+dt] + p_{flip}[t, t+dt] =1$, as it must be. Here ``$nj$'' stands for ``no jump'' (as before), while ``$flip$'' indicates that
the state of the atom performs a \textit{``quantum jump''} from the state parametrized by $\vec{e}(t)$ to its antipode, 
$\vec{e}(t+dt) = -\vec{e}(t)+ \mathcal{O}(dt)$, on the Bloch sphere. Formulae \eqref{flips} follow by writing the state $\Omega\big(\vec{n}(t+dt)\big)$ as a
convex combination of the pure states $$\Omega\big(\pm \vec{n}(t+dt)/\big| \vec{n}(t+dt)\big|\big).$$

Let us suppose that, during a time interval $[t_1, t_2)$, the state of an individual atom, corresponding to a vector $\vec{e}(t)$ on the
Bloch sphere, does \textit{not} perform any quantum jump. For $t_1 \leq t< t_2$, we parametrize $\vec{e}(t)$ by setting
\begin{equation}\label{n}
\vec{e}(t)=\begin{pmatrix} \sqrt{1- e_{3}(t)^{2}} \text{ cos}(t + \varphi_0)\\ \sqrt{1- e_{3}(t)^{2}} \text{ sin}(t + \varphi_0)\\ e_3(t) \end{pmatrix}\,,
\end{equation}
where $\varphi_0$ is a constant determined by the initial vector $\vec{e}(t_1)$. The trigonometric functions on the right side of this
equation describe the precession of $\vec{e}(t)$ around the 3-axis, $\vec{e}_3$, with angular velocity $\omega\equiv 1$.
Re-writing the non-linear evolution equation \eqref{time-dep} of Subsect.~4.1 for the projection $\Pi_t:= \Omega\big(\vec{e}(t)\big)$,
with $\vec{e}(t)$ parametrized as in \eqref{n}, we find a non-linear equation for the 3-component of $\vec{e}(t)$, namely
\begin{equation}\label{nleq}
\dot{e}_3(t)= -\frac{\alpha}{4}\Big(1+e_3(t)\Big) \Big(1- e_3(t)\Big) \Big(2+ e_3(t)\Big)\,.
\end{equation}
Of course, this equation can also be derived directly from equation \eqref{Eq for n} and the state-reduction postulate of the $ETH$ -
Approach, which, in the present case, tells us that the true state of an individual atom is determined by Eqs.~\eqref{purification} and
\eqref{flips}.
The probability that the atom does not perform a quantum jump during the time interval $[t_1, t_2)$ can be determined
with the help of formula \eqref{Feynman-Hellmann} of Subsect.~4.1 or by using Eq.~\eqref{flips}. 

These observations have the following consequences.
\begin{enumerate}
\item{If $\vec{e}(t_1) = -\vec{e}_3$, i.e., $e_{3}(t_1) = -1$, meaning that, at time $t=t_1$, the atom is in the groundstate, then
$$\vec{e}(t)\equiv -\vec{e}_3, \quad \text{and }\quad p_{nj}[t_1, t]\equiv 1, \,\,\forall\,\,t>t_1\,,$$
which says that the atom will remain in its groundstate forever.}
\item{If $\vec{e}(t_1) =  \vec{e}_3$, i.e., $e_3(t_1)= +1$, meaning that, at time $t=t_1$, the atom is in the excited state, then
$\vec{e}(t) = \vec{e}_3, \forall\, t\in[t_1, t_2)$, i.e., the atom remains in the excited state until, at a time $t_2>t_1$, 
it decays into its groudstate. The probability of this trajectory, i.e., the probability of not observing a quantum jump during the
time interval $[t_1, t_2)$, is given by
$$p_{nj}[t_1, t_2] = \text{exp}\big[-\alpha(t_2 - t_1)\big]\,,$$
i.e., an exponential decay law holds. The probability to jump from the excited state to the groundstate in a time interval 
\mbox{$[t, t + dt]$} is given by $ \alpha \, dt\,.$}
\item{If $e_3(t_1)\in (-1,1)$, i.e., at some time $t_1$, the atom is in a coherent superposition of the excited state and the groundstate,
then the vector $\vec{e}(t)$ parametrizing the state of the atom at times $t\in[t_1, t_2)$ precesses around the 3-axis, $\vec{e}_3$,
with angular velocity $\omega=1$, its 3-component $e_3(t)$ solves Eq.~\eqref{nleq}, and the probability $p_{nj}$ for not observing
a quantum jump during the time interval $[t_1,t_2)$ satisfies the differential equation
\begin{equation}\label{eq:pnojump}
\frac{ \ln\Big( p_{nj}[t, t+dt]\Big)}{dt} = -\frac{\alpha}{4}\Big(1+ e_3(t)\Big)^{2}\,,
\end{equation}
whose solution is given by
\begin{align*}
p_{nj}[t_1, t_2]&\equiv p_{nj}[t_1, t_2; \vec{e}(t_1)]=  \text{exp}\big[-\gamma(t_1, t_2)\big]\,, \quad \text{where}\\
&\gamma(t_1, t_2):= \frac{1}{2}\int_{e_3(t_2)}^{e_3(t_1)}\frac{1+\lambda}{(1-\lambda)(2+\lambda)} d\lambda\,.
\end{align*}
}
\item{State trajectories with flips can be treated as in Sect. 4.2, Eq.~\eqref{Poisson measure}, i.e., the probability of observing a 
state trajectory (or ``history''), $\mathfrak{T}$, with initial condition $\Omega_0 = \Omega\big(\vec{e}(0)\big), \big|\vec{e}(0)\big|=1,$ 
and with flips in the time intervals $[t_1, t_1 +dt_1), \dots, [t_n, t_n +dt_n)$, with $0<t_1<\dots t_n<\overline{t}$, is given by 
\begin{align}\label{2-level meas}
W_{\mathfrak{T}}\big[t_1, \dots&, t_n; \vec{e}_0]\prod_{j=1}^{n}dt_j=\\
&= \Big\{\prod_{j=1}^{n} p_{nj}[t_j, t_{j+1}; \vec{e}(t_j)] p_{flip}[t_{j+1}, t_{j+1} + dt_{j+1}; \vec{e}(t_{j+1})]\Big\} p_{nj}[t_n, \overline{t}; \vec{e}(t_n)]\,,
\end{align}
where $p_{flip}[t, t + dt; \vec{e}(t)]= 1- p_{nj}[t, t+dt; \vec{e}(t)]$; see Eq.~\eqref{flips}.
}
\item{These results imply that the state of an individual atom approaches the groundstate, as time $t\rightarrow \infty$.}
\end{enumerate}

\subsection{Evolution of 2-level atoms with photon detection}
In this subsection we specialize the results of Sect.~\ref{Q-Poisson} to 2-level atoms. 
We continue to use the representation of atomic states given in \eqref{Bloch} and the parametrization of the unit vectors $\vec{e}(t)$ 
corresponding to pure states introduced in \eqref{n}. We then re-write equation \eqref{(21')} of Sect.~\ref{Q-Poisson} (see also \eqref{21'}) 
in terms of the unit vectors $\vec{e}(t)$ corresponding to pure states of an individual atom, using the parametrization \eqref{n}.
This yields the equation
\begin{equation}\label{no emission}
\dot{e}_{3}(t) = -\frac{\alpha}{2} \Big(1- e_{3}(t)\Big)\Big(1+ e_{3}(t)\Big)\,.
\end{equation}
Obviously, $e_{3}(t) \equiv \pm 1$ are two constant solutions of this equation. If the initial state $\vec{e}(0)$ is a coherent 
superposition of the excited state and the groundstate then the 3-component, $e_{3}(t)$, of the solution $\vec{e}(t)$ of
\eqref{no emission} is monotone decreasing in time and approaches $-1$, as $t\rightarrow \infty$. The probability for
the atom \textit{not} to spontaneously emit a photon during the time interval $[0, t)$ is given by the following formulae.
\begin{enumerate}
\item[(i)]{If the atom is prepared in its groundstate then it remains in the groundstate forever; i.e., if $e_{3}(0)=-1$ then
$e_{3}(t)\equiv -1, \,\,\forall\,\,t>0,$ and
$$p_{nj}[0, t; \vec{e}(0)] \equiv 1\,.$$
}
\item[(ii)]{If the atom is prepared in its excited state then it remains in the excited state until, at some time $t$, it jumps to its groundstate
to henceforth remain in its groundstate; i.e., if $e_{3}(0)= 1$ then $e_{3}(t')=1, \,\, \forall \,\, t'<t$, for some time $t$, with 
$$p_{nj}[0, t; \vec{e}(0)] = e^{-\alpha t}\,.$$
The probability to jump from the excited state to the groundstate in the time interval \mbox{$[t, t + dt]$} is given by
$$p_{flip}[t, t+dt] = \alpha \, dt,$$
as one infers from the formula for $p_{nj}$ and the fact that $p_{nj}[t, t+dt] + p_{flip}[t,t+dt] =1$.
}
\item[(iii)]{If the atom is prepared in a coherent superposition of its excited state and its groundstate, i.e., for $e_{3}(0)\in(-1,1)$, then
the time-dependence of $e_{3}(t)$ is determined by solving Eq.~\eqref{no emission}, and the probability, $p_{nj}$, 
of not emitting a photon and not jumping to the groundstate is given by
\begin{equation}\label{eq:NoEmi}
p_{nj}[0, t; \vec{e}(0)] = \text{exp}\Big[- \int_{e_{3}(t)}^{e_{3}(0)} \frac{d\lambda}{1-\lambda}\Big]\,.
\end{equation}
When the atom emits a photon recorded by a photomultiplier it simultaneously jumps to its groundstate 
and then stays there forever.}
\end{enumerate}
In Appendices B and C we present some details concerning the derivation of the results in this section.

\section{Conclusions and Outlook}\label{conclusions}

In this paper we have presented results on the phenomenon of fluorescence of atoms coupled to the quantized
electromagnetic field \cite{Pomeau}, which we have derived using the principles of the \mbox{$ETH$ -} Approach 
to Quantum Mechanics applied to simple idealized models. These models describe spontaneous and induced
emission and absorption of photons by static atoms with finitely many internal energy levels in a limiting regime where 
the velocity of light tends to $\infty$. Our analysis has resulted in the derivation of explicit stochastic differential 
equations describing the effective time evolution of states of indvidual atoms. These equations are manifestations 
of a type of stochastic process emerging from the basic principles formulated in the $ETH$ - Approach that we call
\textit{``quantum Poisson process.''} Our results are illustrations of a general theory of \textit{``quantum jumps''} 
emerging from the $ETH$ - Approach, which we expect to play an important role in completing non-relativistic 
Quantum Mechanics to a physically realistic theory free of paradoxes. 

In this paper we have assumed that the radiation field is prepared in its vacuum state. This assumption renders
the analysis of the evolution of the atom and of spontaneous emission of photons by the atom particularly simple; 
(absorption processes of photons do not occur under this assumption). However, our methods enable us to
derive the equations describing the stochastic evolution of such systems under rather general assumptions on 
the preparation of the radiation field and for atoms exhibiting Rabi oscillations. (See \cite{FP} where models with 
discretized time are analyzed.)

It would be of considerable interest to extend the analysis presented in this paper to more realistic models where the
velocity of light is finite, enabling one to calculate relativistic corrections to the results presented in Sects.~\ref{Unraveling Lindblad} - 
\ref{Example}. Discretizing time ($dt>0$) in order to eliminate ultraviolet divergences encountered in quantum
electrodynamics, we have proposed a family of concrete models of this kind; see \cite{FP, FGP}. But it is difficult to 
extract explicit quantitative predictions from these models. The challenge to estimate the size of relativistic corrections 
caused by the finiteness of the velocity of light in a theory of quantum jumps remains to be addressed.

In \cite{FP} we have introduced and studied models with discrete time that describe quantum-mechanical \textit{measurements} 
of physical quantities characteristic of simple physical systems (represented by self-adjoint operators with discrete
spectrum, as usual), and we have  analyzed them using the principles of the $ETH$ - Approach. It is not difficult to let the 
time step in some of those models tend to $0$ and to come up with models similar to the ones studied in Sect.~\ref{Physical systems} 
of this paper. These models can be shown to give rise to Lindblad equations describing the effective time evolution of ensemble states. 
In simple examples these equations are of the form
$$\dot{\Omega}_t = -\frac{i}{\hbar}\big[H, \Omega_t\big] + \sum_{i=1}^{N}\Big[ Q_i V\Omega_t V^{*}Q_i  -
\frac{1}{2}\big\{\Omega_t, V^{*}Q_i V\big\}\Big],$$
where $\big\{Q_1, \dots, Q_N\big\}$ are the orthogonal projections onto eigenspaces of an operator, $X=X^{*}$, representing
the physical quantity to be measured, and the operator $V$ captures certain properties of the measuring device used to measure 
$X$. By ``unraveling'' this equation, using the principles of the $ETH$ - Approach, we are able to come up with a physically
realistic description of measurements of the quantity represented by the operator $X$ -- \textit{without the need to invoke any 
unnatural extra mechanisms or postulates.}  A detailed analysis of such models will appear in another paper. Suffice it to emphasize that, 
in the $ETH$ - Approach to QM, the infamous \textit{``measurement problem''} of Quantum Mechanics evaporates.

For the time being, the $ETH$ - Approach to Quantum Mechanics has only been  illustrated explicitly within the realm of \textit{non-relativistic}
quantum theory. However, an extension of this approach to relativistic quantum theory is feasible; it has been sketched in
\cite{Fr, FGP}. Clearly further efforts will be necessary to make it more precise and to find out possible limitations.
\\

\appendix
\section{Infinitesimal Perturbation Theory}

\noindent
Let $\mathfrak{H}$ be a separable Hilbert space. For the purposes of this paper, we may imagine that $\mathfrak{H}$ 
is finite-dimensional, with dim$(\mathfrak{H})=N<\infty$. We consider two selfadjoint operators, $H_0$ and $H_1:= H_0 + V$, 
where the operator $V$ is interpreted as a perturbation. We assume for simplicity that the eigenvalues, $E_0 < \dots < E_{N-1}$, 
of $H_0$ are non-degenerate and that the corresponding eigenvectors, $\psi_0, \dots, \psi_{N-1}$, are known explicitly; i.e., $H_0$ 
is diagonal in an explicitly known orthonormal basis, $\big\{\psi_j\big\}_{j=0}^{N-1}$, of $\mathfrak{H}$.

We are interested in diagonalizing the operator $H_1$. A standard approach to solving this problem is to make use of analytic
perturbation theory; (see, e.g., \cite{R&S}). In this appendix we sketch an alternative method, called \textit{infinitesimal perturbation theory},
which reduces the problem to solving a system of non-linear ordinary differential equations. 

The key ideas underlying infinitesimal perturbation theory are as follows.\\
We choose a smooth curve, $\big\{H(t) = H(t)^{*}\,\big|\, 0\leq t \leq 1\big\}$, of 
selfadjoint operators, with $H(0)=H_0$ and $H(1)=H_1$, the standard choice being
$$H(t) = H_0 + tV, \quad 0\leq t\leq 1\,.$$
But there are other choices of curves of operators connecting $H_0$ to $H_1$; (for example a curve of operators parametrized 
by a curve in the \textit{complex} $t$-plane connecting $0$ to $1$, see Remark 2, below).
We temporarily assume that the eigenvalues $E_0(t')<\dots < E_{N-1}(t')$ of the operator $H(t')$ 
and the corresponding eigenvectors, $\psi_0(t'), \dots, \psi_{N-1}(t')$, are known for all parameter values 
$t'\in [0,t],$ for some $t<1$. We are interested in determining 
the eigenvalues and eigenvectors of the operator $H(t+dt)$, for sufficiently small values of $dt>0$. We write
\begin{equation}\label{A1}
H(t+dt)= H(t)+ \dot{H}(t) dt + \mathcal{O}(dt^2), \quad \text{with }\quad \dot{H} \equiv \frac{dH}{dt}\,.
\end{equation}
Since we have assumed that the eigenvalues $E_0(t)<\dots < E_{N-1}(t)$ are non-degenrate, i.e., that
$$\underset{j=0,\dots, N-2}{\text{min}} \big[E_{j+1}(t) - E_j(t)\big] =: \Delta(t)>0,$$
the eigenvalues of the operator $H(t+dt)$ and the corresponding eigenvectors can be determined by analytic 
perturbation theory, assuming that $dt$ is small enough -- more precisely $0< dt < \mathcal{O}\big(\Vert \dot{H}(t)\Vert/ \Delta(t)\big)$. 
Let $\Pi_{j}(t):= \big|\psi_j (t)\big>\big< \psi_j(t)\big|$ denote the orthogonal projection onto the $j^{th}$ eigenvector,
$\psi_j(t)$, of the operator $H(t)$. We define an anti-selfadjoint operator, $\mathcal{S}(t)$, by
\begin{equation}\label{A2}
\mathcal{S}(t):= \text{ad}^{-1}_{H(t)}\big[\dot{H}(t)^{od}\big]\equiv \sum_{i\not= j}\frac{\Pi_{i}(t)\,\dot{H}(t)\,\Pi_{j}(t)}{E_i(t)-E_j(t)}\,,
\end{equation}
where the superscript \textit{``od''} stands for ``off-diagonal'', and the sum on the right side ranges over all $i, j = 0, \dots, N-1,$ with $i\not=j$.
Standard formulae of analytic perturbation theory then imply that
\begin{align}\label{A3}
\begin{split}
&E_{j}(t+dt) = E_{j}(t) + \text{Tr} \big(\Pi_{j}(t) \,\dot{H}(t)\big) dt + \mathcal{O}(dt^2), \quad \text{and} \\
&\Pi_{j}(t+dt) = \Pi_{j}(t) - \big[\mathcal{S}(t), \Pi_{j}(t)\big]dt + \mathcal{O}(dt^{2})\,.
\end{split}
\end{align}
Letting $dt\rightarrow 0$, we find the following system of non-linear ordinary differential equations
for the eigenvalues $E_{j}(t)$ and eigenprojections $\Pi_{j}(t)$, $j=0, \dots, N-1$:
\begin{align}\label{A4}
\begin{split}
&\dot{E}_{j}(t)= \text{Tr}\big( \Pi_{j}(t)\,\dot{H}(t)\big)\, \quad \text{(Feynman-Hellmann theorem)}\\
&\dot{\Pi}_{j}(t)=-\big[\mathcal{S}(t), \Pi_{j}(t)\big]\,, \quad \text{for }\,\,j=0, \dots, N-1\,.
\end{split}
\end{align}
Plugging expression \eqref{A2} for $\mathcal{S}(t)$ into the right side of the second equation in \eqref{A4}, we see 
that \eqref{A4} is a closed system of non-linear, first-order ordinary differential equations.\\

\textit{Remarks:}
\begin{enumerate}
\item{The method sketched here can be generalized to deal with degenerate eigenvalues.}
\item{If $N< \infty$ one need not assume that the operators $H(t)$ be selfadjoint. The curve
$\big\{ H(t)\,\big|\, 0\leq t \leq 1\big\}$ can make excursions into the space of \textit{non-selfadjoint} matrices. 
This flexibility is useful in the analysis of the eigenvalue spectrum and the eigenvectors of the operator$H(1)=H_0 + V$
in situations where the standard curve \mbox{$H(t)=H_0 + tV, 0\leq t\leq 1$,} passes through operators $H(t_{*})$ where eigenvalues
degenerate (corresponding, e.g., to eigenvalue crossings at $t=t_{*}$), so that the gap $\Delta(t)$ introduced above
vanishes at $t= t_{*}$, which makes ordinary analytic perturbation theory diverge, as $t$ approaches $t_{*}$.}
\item{One may want to study a \textit{variational problem} to determine an optimal curve \mbox{$\{H(t)\}_{0\leq t\leq 1}$} 
connecting $H_0$ to $H_1$ by minimizing a certain \textit{``cost functional''} related, e.g., to the computational complexity 
encountered in solving the system of equations in \eqref{A4}.}
\end{enumerate}
\vspace{0.3cm}

\section{Proofs of Equations \eqref{nleq} and \eqref{eq:pnojump} of Section \ref{Example}}

We begin by reformulating the problems addressed in Sect.~\ref{Example} in a convenient form.

We first observe that the density matrix, $\Omega_{\downarrow}$, defined by
\begin{align*}
    \Omega_{\downarrow}:=\left(
    \begin{array}{cc}
        0 & 0 \\
        0 & 1
    \end{array}
    \right),
\end{align*} is a static solution to the Lindblad equation \eqref{emission}, because $\mathcal{L}_{\alpha}\big[\Omega_{\downarrow}\big]=0$.
Moreover, for an arbitrary density matrix $\Omega$,  $\Omega - \Omega_{\downarrow}$ is a Hermitian $2\times2$ matrix of trace 0 
and hence can be written in the form
\begin{align}\label{Omega}
    \Omega - \Omega_{\downarrow}=a \sigma_1+b\sigma_2+c\sigma_3, 
\end{align} for some real numbers $a,\ b,$ and $\ c$.

Thus, to study the Lindblad evolution 
\begin{equation}\label{L-eq}
\Omega_t:=e^{t\mathcal{L}_{\alpha}}\big[\Omega\big]=e^{t\mathcal{L}_{\alpha}}\big[\Omega-\Omega_{\downarrow}\big]+\Omega_{\downarrow},
\end{equation}
it suffices to consider $e^{t\mathcal{L}_{\alpha}}\big[a \sigma_1+b\sigma_2+c\sigma_3\big],$ with $a,\ b,\ c\in \mathbb{R}.$

It is convenient to find an explicit matrix representation for $\text{exp}(\mathfrak{L}_{\alpha})$ acting on a linear vector space, $\mathcal{V}$, 
over $\mathbb{R}$, in order to make standard techniques available; $\mathcal{V}$ is defined by
\begin{align}
    \mathcal{V}:=\Big\{ a \sigma_1+b\sigma_2+c\sigma_3\ \Big|\ a,\ b,\ c\in \mathbb{R} \Big\}.
\end{align} 
Our choice of $\mathcal{L}_{\alpha}: \ \mathcal{V}\rightarrow \mathcal{V}$ in Eq.~\eqref{emission} implies that
\begin{align}\label{eq:LoverV}
\begin{split}
   \mathcal{L}_{\alpha}\big[a\sigma_1+b\sigma_2+c\sigma_3\big]
    =(-b-\frac{\alpha }{2}a)\sigma_1+(a-\frac{\alpha }{2}b)\sigma_2-\alpha c\sigma_3.
\end{split}
\end{align}
Hence $\mathcal{L}_{\alpha}$ can be written as a \textit{real} $3\times 3$ matrix, $\mathcal{M}$, on $\mathbb{R}^ {3}$ given by
\begin{align}
    \mathcal{M}=\left(
    \begin{array}{ccc}
        -\frac{\alpha }{2} & -1 & 0 \\
        1 & -\frac{\alpha }{2} & 0\\
         0 & 0 & -\alpha 
    \end{array}
    \right).
\end{align} 
Defining $\vec{n}_t := e^{t \mathcal{M}}\vec{n}_0,$ for an arbitrary $\vec{n}_0 \in \mathbb{R}^{3}$, we have that
\begin{align}\label{Evo}
    e^{t \mathcal{L}_{\alpha}}\big[\vec{n}_0\cdot \vec{\sigma}\big]=\vec{n}_{t}\cdot \vec{\sigma}.
\end{align}
To find an explicit formula for $\vec{n}_t$, we must study the eigenvalues and eigenvectors of the matrix $\mathcal{M}$:
its eigenvalues are given by
$$-\frac{\alpha }{2}+i,\ -\frac{\alpha }{2}-i,\  -\alpha ,$$
and the corresponding eigenvectors by
$$
\vec{e}_1-i\vec{e}_2,\ \vec{e}_1+i\vec{e}_2\,,\ 
  \vec{e}_3.
$$ 
where $\{\vec{e}_k\ |\ k=1,2,3\}$ is the standard basis in $\mathbb{R}^3$.

We are now ready to compute $e^{t\mathcal{M}} \vec{e}_k, k=1,2,3$.
\begin{align*}
    e^{t\mathcal{M}}\vec{e}_1=&\frac{1}{2} e^{tM}\Big\{(\vec{e}_1-i\vec{e}_2)+(\vec{e}_1+i\vec{e}_2)\Big\}
    = e^{-\frac{\alpha }{2}t} \Big(\cos(t)\vec{e}_1+\sin(t)\vec{e}_2\Big),
\end{align*} and similarly
\begin{align*}
        e^{t\mathcal{M}}\vec{e}_2=\frac{1}{2}i e^{tM}\Big\{(\vec{e}_1-i\vec{e}_2)-(\vec{e}_1+i\vec{e}_2)\Big\}
    =e^{-\frac{\alpha }{2}t} \Big(\cos(t)\vec{e}_2-\sin(t)\vec{e}_1\Big).
\end{align*} 
Thus, by \eqref{Evo},
\begin{align}
\begin{split}
    e^{t\mathcal{L}_{\alpha}}\big[\sigma_1\big] =&e^{-\frac{\alpha }{2}t} \Big(\cos(t)\sigma_1+\sin(t)\sigma_2\Big),\\
    e^{t\mathcal{L}_{\alpha}}\big[\sigma_2\big] =&e^{-\frac{\alpha }{2}t} \Big(\cos(t)\sigma_2-\sin(t)\sigma_1\Big).
\end{split}
\end{align}
Since $e^{t\mathcal{M}}\vec{e}_3=e^{-\alpha t}\vec{e}_3,$ we have that
\begin{align}
    e^{t \mathcal{L}_{\alpha}}\big[\sigma_3\big]=e^{-\alpha t}\sigma_3.
\end{align}
In conclusion,
\begin{align}\label{eq:evoabc}
\begin{split}
    &e^{t\mathcal{L}_{\alpha} }\big[a\sigma_1+b\sigma_2+c\sigma_3\big]
    =\left(
    \begin{array}{cc}
        ce^{-\alpha t}   & e^{-\frac{\alpha }{2}t} \sqrt{a^2+b^2}\ e^{-i(t+\gamma)}  \\
          e^{-\frac{\alpha }{2}t} \sqrt{a^2+b^2}\ e^{i(t+\gamma)} & -ce^{-\alpha t}
    \end{array}
    \right),
\end{split}
\end{align} where $\gamma\in [0,2\pi)$ is determined by requiring that $\cos(\gamma)=\frac{a}{\sqrt{a^2+b^2}}$ and $\sin(\gamma)=\frac{b}{\sqrt{a^2+b^2}},$ for $a^2 + b^2 \not=0$, and we use the well-known identities
\begin{align*}
    \cos(t)\cos(\gamma)-\sin(t)\sin(\gamma)=&\cos(t+\gamma),\\
    \sin(t)\cos(\gamma)+\cos(t)\sin(\gamma)=&\sin(t+\gamma).
\end{align*}
(If $a^2+b^2=0$ we set $\gamma=0$; the value of $\gamma$ does not matter, because $\sqrt{a^2+b^2} e^{i\gamma}=0$.)

Returning to the Lindblad equation \eqref{L-eq} for the density matrix, with initial conditions at time $t=\underline{t}$ given by 
\begin{align}\label{eq:abc}
\Omega_{0}=\Omega_{\downarrow}+a\sigma_1+b\sigma_2+c\sigma_3,
\end{align}
we find that
\begin{align}\label{eq:exMt}
    e^{(t-\underline{t}) \mathcal{L}_{\alpha}} \big[\Omega_0\big]=\frac{1}{2}\Big(1+\vec{n}(t)\cdot \vec\sigma\Big)\,,
\end{align} 
where $\vec{n}(t)$ is given by
\begin{align}
    \vec{n}(t):=\left(
   \begin{array}{cc}
      2e^{-\frac{\alpha }{2}(t-\underline{t})} \sqrt{a^2+b^2} \cos(t-\underline{t}+\gamma)  \\
      2 e^{-\frac{\alpha }{2}(t-\underline{t})} \sqrt{a^2+b^2}\sin(t-\underline{t}+\gamma)  \\
      2c e^{-\alpha (t-\underline{t})}-1  
   \end{array}
    \right).
\end{align} 
We choose the initial condition $\Omega_0$ to be given by a pure state 
\begin{align}
    \Omega_{0}=\frac{1}{2}\Big(1+\vec{e}_0\cdot\vec{\sigma}\Big)\,,\qquad |\vec{e}_0| = 1\,,
\end{align} 
with
$$\vec{e}_0= \begin{pmatrix} e_1\\e_2\\e_3 \end{pmatrix}= \left(
\begin{array}{c}
     \sqrt{1-e_3^2} \cos(\gamma)  \\
     \sqrt{1-e_3^2} \sin(\gamma) \\
     e_3
\end{array}
\right)$$ for some $e_3 \in [-1,1]$ and $\gamma\in [0,2\pi).$ The constants $a,$ $b$ and $c$ in \eqref{eq:abc} are thus given by
\begin{align*}
    a=&\frac{1}{2}\sqrt{1-e_3^2} \cos(\gamma),\\
    b=&\frac{1}{2}\sqrt{1-e_3^2} \sin(\gamma),\\
    c=&\frac{1}{2}\Big(1+e_3\Big).
\end{align*}
By \eqref{Evo},
\begin{align}
   \Omega_t := e^{(t-\underline{t})\mathfrak{L}_{\alpha}}\big[\Omega_{0}\big]=\frac{1}{2} \Big(1+\vec{n}(t)\cdot \vec\sigma \Big)
\end{align} 
with 
\begin{equation}\label{vector}
\vec{n}(t)=\left(
\begin{array}{c}
      e^{-\frac{\alpha }{2}(t-\underline{t})}\sqrt{1-e_3^2} \cos(\gamma+t-\underline{t}) \\
      e^{-\frac{\alpha }{2}(t-\underline{t})}\sqrt{1-e_3^2} \sin(\gamma+t-\underline{t})\\
      \Big(1+e_3\Big)e^{-\alpha (t-\underline{t})}-1
\end{array}
\right).
\end{equation}
Equations \eqref{Evo} and \eqref{vector} determine the time evolution of ensemble states.
We observe that, for $t>\underline{t}$,
\begin{align}\label{def:qt1}
q(t):=|\vec{n}(t)|<1, \quad \text{i.e. }\,\,\Omega_t \,\,\text{is a } mixed \text{ state}\,,
\end{align}
unless $e_3=-1$ (i.e., the atom is in its ground-state), and $\lim_{t\rightarrow \underline{t}} q(t)=1$. The state $\Omega_t$ 
can be decomposed into a convex combination of pure states
\begin{align}\label{97}
    \Omega_t=\frac{1+q(t)}{4} \Big(1+\vec{e}(t)\cdot \vec\sigma \Big)
+\frac{1-q(t)}{4}\Big(1-\vec{e}(t)\cdot \vec\sigma \Big)\,,
\end{align}
where $\vec{e}(t):=\frac{\vec{n}(t)}{|\vec{n}(t)|}.$

Following the general theory developed in Sect.~\ref{Unraveling Lindblad}, we derive the differential equation given in \eqref{nleq}. 
We assume that 
$$\Omega_t= \frac{1}{2}\big(\mathbf{1} + \vec{e}(t)\cdot \vec{\sigma}\big), \qquad |\vec{e}(t)| = 1\,,$$
is a pure state, and then define $\Omega_{t+dt}, dt>0,$ by 
\begin{equation}\label{Inf-evo}
\Omega_{t+dt} \equiv \frac{1}{2}\Big(\mathbf{1}+ \vec{n}(t+dt)\cdot \vec{\sigma}\Big):= e^{dt \mathfrak{L}_{\alpha}}\big[\Omega_t\big] = \Omega_t + \mathfrak{L}_{\alpha}\big[\Omega_t\big]\,dt + \mathcal{O}(dt^2)\,.
\end{equation}
We are interested in the third component, $e_3(t+dt)$, of the vector $\vec{e}(t+dt)=\vec{n}(t+dt)/|\vec{n}(t+dt)|$, which is related to the third component, $n_3(t+dt)$, of $\vec{n}(t+dt)$ by a correction of order $\mathcal{O}(dt^2)$, with $0\leq dt \ll 1$. Equation \eqref{vector} yields
\begin{align}
\begin{split}
e_3(t+dt)=&\frac{\Big(1+e_3(t)\Big)e^{-\alpha dt} -1}{\sqrt{e^{-\alpha dt}\Big(1-e_3^2(t)\Big)+\Big(\big(1+e_3(t)\big)e^{-\alpha dt}-1 \Big)^2}}
     +\mathcal{O}(dt^2)\\
    =&e_3(t)-\frac{\alpha}{4}\Big(1+e_3(t)\Big) \Big(1- e_3(t)\Big) \Big(2+ e_3(t)\Big)dt+\mathcal{O}(dt^2),
\end{split}
\end{align} where we have used the Taylor expansion $e^{-\alpha dt}=1-\alpha dt +\mathcal{O}(dt^2).$
Taking $e_3(t)$ to the left side and letting $dt$ tend to 0, we obtain the differential equation
\begin{align}
    \dot{e}_{3}(t)\equiv \frac{d}{dt}e_3(t)=-\frac{\alpha}{4}\Big(1+e_3(t)\Big) \Big(1- e_3(t)\Big) \Big(2+ e_3(t)\Big)\,,
\end{align} 
which is equation \eqref{nleq}.

The quantity $\frac{1+q(t)}{2}$, see \eqref{def:qt1} and \eqref{97}, is related to the ``no-jump'' probability $p_{nj}[t, t+dt]$ in \eqref{eq:pnojump} by the identity
\begin{align*}
p_{nj}[t, t+dt] = \frac{1+ |\vec{n}(t+dt)|}{2} +\mathcal{O}(dt^2)
\end{align*}
Consequently,
\begin{align}
    \frac{\ln \Big(p_{nj}[t, t+dt]\Big)}{dt}=\frac{\alpha}{4}\Big(1+e_3(t)\Big)^{2},
\end{align}
which is equation \eqref{eq:pnojump}.

\section{Proofs of Equations \eqref{no emission} and \eqref{eq:NoEmi} of Section \ref{Example}}\label{sec:noemissions}
Since the proof of \eqref{no emission} is easier than the one of \eqref{nleq}, we will only sketch it.

For the two-level atom, the operator $\mathfrak{L}'_{\alpha}$, defined in \eqref{21'} (see also \eqref{(21')}), has two parts
\begin{align}
\mathfrak{L}'_{\alpha}=K_1+K_2,
\end{align} where $K_1$ and $K_2$ are defined by
\begin{align*}
    K_1\big(\Sigma\big):=&- i\ ad_{\frac{1}{2}\sigma_3}\big(\Sigma\big),\\
    K_2\big(\Sigma \big):=&-\frac{1}{2}\alpha \Big\{\Sigma,\ \sigma_{+}\sigma_{-}\Big\},
\end{align*}
and $\Sigma$ is an arbitrary Hermitian matrix.

We note that $K_1$ and $K_2$ commute with each other, which enables us to solve equation \eqref{21'} explicitly.
If $\Omega_t$ is a pure state it is of the form
\begin{align}\label{eq:mtc}
\Omega_t=\left(
\begin{array}{cc}
1-c & a e^{-i\gamma}\\
ae^{i\gamma} &c
\end{array}
\right),
\end{align} for some $\gamma\in [0,2\pi)$, $c\in [0,1]$, and $a\geq 0$ satisfying 
\begin{align*}
    a^2=c(1-c).
\end{align*}
(We suppress the dependence on time $t$ of $a$, $c$ and $\gamma$ in our notation.) Since $K_1$ and $K_2$ commute, 
$$\Omega_{t+dt}:=e^{\mathfrak{L}'_{\alpha}dt}\big[\Omega_t\big]=e^{K_2 dt}e^{K_1 dt}\big[\Omega_t\big]$$ 
is given by
\begin{align}\label{eq:ntdt}
    \Omega_{t+dt}=\left(
    \begin{array}{cc}
      e^{-\alpha dt}(1-c)   & e^{-\frac{\alpha dt}{2}} a e^{-i(\gamma+dt)} \\
       e^{-\frac{\alpha dt}{2}} a e^{i(\gamma+dt)}  & c
    \end{array}
    \right).
\end{align}

We observe that the matrix $\Omega_{t+dt}\geq 0$ is of rank one, for its determinant vanishes, and when $0< dt\ll 1,$ 
\begin{align*}
    0<\text{Tr}\big(\Omega_{t+dt}\big)< 1.
\end{align*} 
Thus, $\Pi_{t+dt}:=\frac{1}{q(dt)} \Omega_{t+dt}$ is a pure state, with 
\begin{align}\label{def:qt}
    q(dt):=\text{Tr}\big(\Omega_{t+dt}\big)=e^{-\alpha dt}(1-c)+c.
\end{align}
Since $\Pi_t \equiv \Omega_t$ is assumed to be a pure state, it is of the form
\begin{align*}
    \Omega_t=&\frac{1}{2}\Big(1+\vec{e}(t)\cdot \vec{
\sigma}\Big),
\end{align*} 
for some real unit vector $\vec{e}(t)$, which we write as
\begin{align}\label{106}
    \vec{e}(t)=\left(
    \begin{array}{c}
          \sqrt{1-e_3^2(t)}\cos(\gamma) \\
          \sqrt{1-e_3^2(t)}\sin(\gamma)\\
          e_3(t)
    \end{array}
    \right).
\end{align}
and, in the parametrization of \eqref{eq:mtc},
\begin{align}\label{eq:tn3t}
\begin{split}
    c=&\frac{1}{2}\Big(1-e_{3}(t)\Big),\quad \text{and}\\
    a=&\sqrt{1-e_3^2(t)}.
\end{split}
\end{align}

Since $\Pi_{t+dt}:= \frac{1}{q(dt)} \Omega_{t+dt}$ is a pure state, there is a real unit vector $\vec{e}(t+dt)$ such that 
\begin{align}
\frac{1}{q(dt)} \Omega_{t+dt}=&\frac{1}{2}\Big(1+\vec{e}(t+dt)\cdot \vec{\sigma}\Big).
\end{align}  
By \eqref{106},
\begin{align}
\vec{e}(t+dt)=\left(
    \begin{array}{c}
          \sqrt{1-e_3^2(t+dt)}\cos(\gamma+dt) \\
          \sqrt{1-e_3^2(t+dt)}\sin(\gamma+dt)\\
          e_3(t+dt)
    \end{array}
    \right),
\end{align} and \eqref{eq:ntdt} and \eqref{def:qt} imply that
\begin{align}
e_3(t+dt)&:=1-\frac{2c}{e^{-\alpha dt}(1-c)+c}.
\end{align}
Using \eqref{eq:tn3t} to replace $c$ and expanding in $dt$ to first order, we find
\begin{align}\label{eq:expand1}
\begin{split}
e_3(t+dt)&=1-\frac{2(1-e_3(t))}{e^{-\alpha dt}(1+e_3(t))+(1-e_3(t))}\\
&=e_3(t)-\frac{\alpha }{2} \ \Big(1+e_3(t)\Big)\Big(1-e_3(t)\Big)\ 
 dt+\mathcal{O}(dt^2).
\end{split}
\end{align}
By \eqref{def:qt}
\begin{align}\label{eq:qdt}
    q(dt)=e^{-\alpha dt}\frac{1+e_3(t)}{2}+\frac{1-e_3(t)}{2}=1-\frac{\alpha }{2}(1+e_3(t)) dt+\mathcal{O}((dt)^2).
\end{align}

We are ready to derive equations \eqref{no emission} and \eqref{eq:NoEmi}. By \eqref{def:qt}

\begin{align}
p_{nj}[t, t+dt] = q(dt)+\mathcal{O}((dt)^2),
\end{align}
which, together with \eqref{eq:expand1} and \eqref{eq:qdt}, implies that
\begin{align}
   \dot{e}_3(t) =-\frac{\alpha}{2}\Big(1-e_3(t)\Big)\Big(1+e_3(t)\Big),
\end{align} 
and 
\begin{align}
    \frac{\ln \Big( p_{nj}[t, t+dt]\Big) }{dt}=-\frac{\alpha}{2}\Big(1+e_3(t)\Big) , 
\end{align} 
which implies equation \eqref{eq:NoEmi}.

This completes our proofs.


\begin{thebibliography}{}
\bibitem{...} In the literature one finds many papers on stochastic differential equations presumed to describe 
time evolution (under ``continuous measurements'') in quantum mechanics. We find it quite impossible to determine 
which of these papers might be relevant for the material presented in our paper.
\end{thebibliography}
\end{document}